\pdfoutput=1 

\documentclass[aip, reprint, floats, floatfix]{revtex4-1}

\usepackage{graphicx}
\usepackage{dcolumn}

\usepackage[table]{xcolor}
\usepackage{ctable}
\usepackage{enumitem} 

\usepackage{amsfonts} 
\usepackage{amsmath} 
\usepackage{amssymb} 
\usepackage{bm}

\usepackage{nameref} 

\usepackage{natbib}

\begin{document}

\title[Emergence and maintenance of modularity]
{Emergence and maintenance of modularity in neural networks with Hebbian and anti-Hebbian inhibitory STDP}

\author{Rapha\"el Bergoin}  \email{raphael.bergoin@gmail.com}
\affiliation{ETIS, UMR 8051, ENSEA, CY Cergy Paris Universit\'e, CNRS, 6 Av. du Ponceau, 95000 Cergy-Pontoise, France.}
\affiliation{Center for Brain and Cognition, Pompeu Fabra University, Barcelona, Spain.}
\affiliation{Department of Information and Communication Technologies, Pompeu Fabra University, Barcelona, Spain.}
\affiliation{Institute of Neural Information Processing, Center for Molecular Neurobiology (ZMNH), University Medical Center Hamburg-Eppendorf (UKE), 20251 Hamburg, Germany.}

\author{Alessandro Torcini} 
\affiliation{Laboratoire de Physique Th\'eorique et Mod\'elisation, UMR 8089, CY Cergy Paris Universit\'e, CNRS, 2 Av. Adolphe Chauvin, 95032 Cergy-Pontoise, France.}

\author{Gustavo Deco} 
\affiliation{Center for Brain and Cognition, Pompeu Fabra University, Barcelona, Spain.}
\affiliation{Department of Information and Communication Technologies, Pompeu Fabra University, Barcelona, Spain.} 
\affiliation{Instituci\`{o} Catalana de Recerca i Estudis Avan\c{c}ats (ICREA), Passeig Lluis Companys 23, 08010 Barcelona, Spain.}

\author{Mathias Quoy} 
\affiliation{ETIS, UMR 8051, ENSEA, CY Cergy Paris Universit\'e, CNRS, 6 Av. du Ponceau, 95000 Cergy-Pontoise, France.}
\affiliation{IPAL, CNRS, 1 Fusionopolis Way \#21-01 Connexis (South Tower), Singapore 138632, Singapore.}

\author{Gorka Zamora-L\'opez} 
\affiliation{Center for Brain and Cognition, Pompeu Fabra University, Barcelona, Spain.}
\affiliation{Department of Information and Communication Technologies, Pompeu Fabra University, Barcelona, Spain.} 

\date{\today}

\begin{abstract} 	
The modular and hierarchical organization of the brain is believed to support the coexistence of segregated (specialization) and integrated (binding) information processes. A relevant question is yet to understand how such architecture naturally emerges and is sustained over time, given the plastic nature of the brain's wiring. Following evidences that the sensory cortices organize into assemblies under selective stimuli, it has been shown that stable neuronal assemblies can emerge due to targeted stimulation, embedding various forms of synaptic plasticity in presence of homeostatic and/or control mechanisms. 
Here, we show that simple spike-timing-dependent plasticity (STDP) rules, based only on pre- and post-synaptic spike times, can also lead to the stable encoding of memories in the absence of any control mechanism.
We develop a model of spiking neurons, trained by stimuli targeting different sub-populations. The model satisfies some biologically plausible features: (i) it contains excitatory and inhibitory neurons with Hebbian and anti-Hebbian STDP; (ii) neither the neuronal activity nor the synaptic weights are frozen after the learning phase. Instead, the neurons are allowed to fire spontaneously while synaptic plasticity remains active.
We find that only the combination of two inhibitory STDP sub-populations allows for the formation of stable  modules in the network, with each sub-population playing a distinctive role.
The Hebbian sub-population controls for the firing activity, while the anti-Hebbian neurons promote pattern selectivity. After the learning phase, the network settles into an asynchronous irregular resting-state. This post-learning activity is associated with spontaneous memory recalls which turn out to be fundamental for the long-term consolidation of the learned memories. Due to its simplicity, the introduced model can represent a test-bed for further investigations on the role played by STDP on memory storing and maintenance.
\end{abstract}

\maketitle

\clearpage
\newpage
\mbox{~}
\clearpage
\newpage
\section*{Introduction}
\label{sec:introduction} 

The brain's connectivity follows a modular and a hierarchical organization at different spatial and functional scales~\cite{zamora2010cortical,varshney2011elegans,sporns2013network,goulas2020hierarchy,suarez2022taxonomy,lin2024drosophila}. From an operative point of view, this type of architecture is suggested to facilitate the coexistence of segregation and integration of information~\cite{sporns2001classes,shanahan2010broadcast,zamora2011exploring}: neuronal circuits or brain regions associated to a specific function are densely connected with each other~\cite{scannell1995analysis,hilgetag2000anatomical,meunier2010modular}, while long-range connections and network hubs allow for the integration (or binding) of different information~\cite{zamora2011exploring,senden2014rich,bertolero2015modular}. 
A crucial open question in brain connectivity is to understand how such modular and hierarchical organization naturally emerges as a consequence of the functional needs of the nervous system.

This question has been addressed from various standpoints.  
On the one hand, it has been shown that spontaneous neural activity might be sufficient to shape network structure~\cite{rubinov2009,damicelli2019,manz2023,li2024review}. The underlying idea is that neural dynamics split brain connectivity into different populations via Hebbian-like adaptation mechanisms that reinforce links between partially synchronised neurons. Such mechanisms may take place, for example, during early development when cortical areas become organized in the absence of sensory stimuli~\cite{kirkby2013review}. The resulting complex connectivity---in turn---supports a combination of time-scales and this facilitates the onset of self-sustained metastable neural activity~\cite{rubinov2009,shanahan2010metastable}, e.g., the spontaneous switching of activity across neural populations~\cite{shanahan2007spiking,schaub2015}. 

On the other hand, the relation between learning and network organization is inherently associated with the notions of semantic memory and input selectivity. Working memory is characterised by synaptic changes leading to groups of neurons (assemblies or engrams) to sustain higher frequencies for a few seconds after stimulus presentation~\cite{fuster1973,miyashita2000review}. These changes also facilitate recognition if the stimulus is presented again shortly after. 
Sensory cortices contain neurons selectively firing for different features of the inputs, forming differentiated groups of neurons (assemblies) related to receptive fields~\cite{hubel1962}. Following these observations, computational models of spiking neurons have been proposed to investigate how memories could be imprinted into the neuronal architecture~\cite{del2001,del2003,amit2003,shanahan2007spiking,morrison2007,clopath2010,litwin2014,zenke2015,ocker2019,fauth2019,miehl2022formation,manz2023,bergoin2023inhibitory,bergoin2023phd,yang2024}. Starting from a random connectivity, their goal is to reproduce the formation of neuronal assemblies in response to various external stimuli (or memories), mediated by synaptic plasticity.

This process comes with several challenges and questions: 
(i) the correct formation of neural assemblies (modules) in the network architecture that are consistent with the learned stimuli, 
(ii) the long-term stability of these structures when the entrainment is finished, and 
(iii) the stability of the resulting network dynamics. 
The mechanism for long-term memory maintenance and consolidation has been related to spontaneous memory recalls (or retrievals). During wake, short and random events of partial synchronous activation of neural sub-networks are related both to the spontaneous recalls~\cite{gu2016dynamics} and to the  consolidation~\cite{vogels2011inhibitory,carrillo2016imprinting,lagzi2021assembly} of learned memories.  Another important aspect is then to clarify how the typical cortical dynamics, that is asynchronous and irregular, can coexist with these spontaneous recalls~\cite{vreeswijk1996chaos,brunel2000dynamics}.

A crucial aspect for the success of the proposed neural networks is the selection of a model for the plasticity.
The exact mechanisms underlying synaptic plasticity have been long debated but the pioneering experiments by Bi \& Poo~\cite{bi1998synaptic,bi2001} on the hippocampus revealed that the plastic changes are driven by a temporally asymmetric form of Hebbian learning, induced by tight temporal correlations between the spikes of pre- and postsynaptic neurons. This phenomenon has been termed as spike-timing-dependent plasticity (STDP)~\cite{stdp,caporale2008,mikkelsen2013}. 
However, previous attempts to model the storage of memory items via stimulations of selected neurons in random balanced networks have led to unstable behaviours impeding the formation of stimulus induced structures~\cite{morrison2007}. To avoid these pathological behaviours other models have been proposed which, instead, combine different types of voltage- and rate-dependent STDP, often in presence of homeostatic or other control mechanisms to prevent instabilities. In particular, in Refs.~\cite{litwin2014,yang2024} neuronal assemblies were successfully formed which reflected the stored memory items, and their long-term maintenance was mediated by spontaneous recalls of the stored patterns. The main difference among these two studies is how the STDP rules were implemented. 
In Litwin-Kumar \& Doiron~\cite{litwin2014} a voltage-based STDP rule~\cite{clopath2010} for the excitatory-excitatory synapses was employed to achieve a stable evolution, together with additional homeostatic mechanisms based on inhibitory plasticity and on non-local renormalization of the excitatory-excitatory synaptic weights. 
In Yang \& La Camera~\cite{yang2024} plastic synapses were considered only among excitatory neurons and no further homeostatic mechanisms. The STDP rule is similar to the one in Ref.~\cite{litwin2014} where a voltage-sensitive variable---depending on the post-synaptic membrane potential---is compared to a threshold in order to determine whether the synapse enters a long-term potentiation or depression state. The main difference is that the thresholds in Yang \& La Camera~\cite{yang2024} evolve dynamically as in the BCM rule~\cite{bienenstock1982}, adapting to the single neuron post-synaptic activity in order to avoid instabilities in the dynamics.

In the present paper, we consider simpler STDP mechanisms based only on the local information associated with pre- and post-synaptic spike times analogous to the one discovered by Bi and Poo~\cite{bi1998synaptic,bi2001}. Contrary to the results in Morrison et al.~\cite{morrison2007} our network model successfully encodes and sustains stimulus-driven assemblies in the absence of any additional control mechanism. This is achieved by accounting for inhibitory STDP mechanisms alongside the excitatory one.
Recent empirical studies have underlined the variety of functional roles played by different classes of interneurons with, for example, parvalbumin-expressing (PV) interneurons in the mouse being subject to symmetric Hebbian STDP and somatostatin-expressing (SOM) ones following asymmetric Hebbian STDP~\cite{lagzi2021assembly,lamsa2005hebbian,lamsa2007anti}. 
Accordingly, our model considers two inhibitory neuronal sub-populations which self-organise during the training phase into two differential functional roles: one sub-population subjected to Hebbian STDP (which controls the firing activity) and another population following anti-Hebbian STDP (which mediates for memory selectivity). 

After the learning phase the model settles into an asynchronous irregular state, as typically observed during \emph{in-vivo} recordings of the brain activity at rest~\cite{softky1993}. This post-learning activity is characterized by the occurrence of transient events of partial synchrony, associated to the learned items as in Refs.~\cite{litwin2014,zenke2015,ocker2019,yang2024}. These spontaneous memory recalls are crucial for the long-term consolidation of the stored memories by promoting the reinforcement of the underlying connectivity. Given that inhibitory plasticity takes an active role in the model, the memory capacity of the network is controlled by the number of inhibitory neurons. Finally, we also demonstrate the emergence of hub neurons displaying mixed selectivity, by training the network to overlapping memory items~\cite{fauth2019,bergoin2023inhibitory,bergoin2023phd,manz2023,yang2024}. These excitatory hub neurons represent the seeds for hierarchical organization and integration.

\section*{Emergence of modular connectivity driven by learning to selective stimuli} 	

We consider networks of excitatory and inhibitory Quadratic Integrate-and-Fire (QIF)~\cite{ermentrout1986} neurons, pulse coupled via exponentially decaying post-synaptic potentials and in the presence of STDP. The connections involving pre-synaptic excitatory neurons are subject to Hebbian STDP, while those involving pre-synaptic inhibitory neurons can be subject to two types of STDP: either Hebbian or anti-Hebbian.

Firstly, we will investigate the necessary conditions for the emergence of modular assemblies induced by learning selective stimuli. For this purpose, we will analyze the role played by Hebbian and anti-Hebbian plasticity applied to inhibitory neurons in a network subject to two external stimuli. Secondly, we will investigate the role of spontaneous recalls in order to consolidate and maintain both the learned memory items and the underlying modular connectivity. To clarify this aspect we will perturb the synaptic connectivity matrix and we will examine to which extent spontaneous recalls are able to regenerate the original structure induced by the training. The robustness of this scenario will be validated by considering larger system sizes and random networks. Finally, we will generalize the model to account for an increasing number of stored memories and in the presence of overlapping assemblies, thus showing that more complex architectures can develop and be maintained over time.

\begin{figure*}[p]
	\centering
	\includegraphics[width=0.75\textwidth]{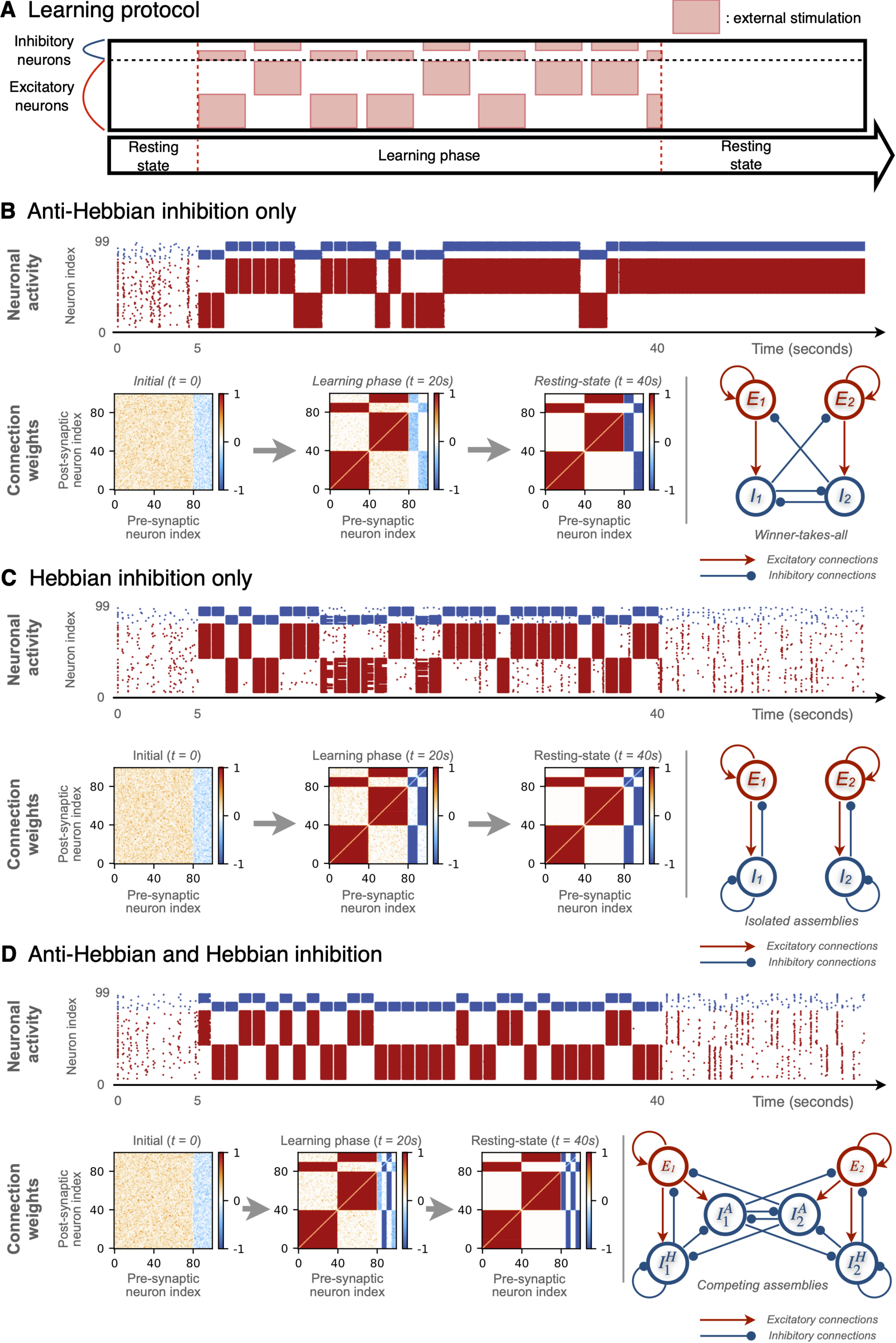}
	\caption{
	{\bf Emergence of modular connectivity by learning two selective stimuli in spiking networks.}
	{\bf (A)} Experimental protocol consists of the stimulation of two non-overlapping neuronal populations of QIF neurons with plastic synapses. Networks are made of $80$\% of excitatory and $20$\% of inhibitory neurons. Stimuli are presented in temporal alternation.
	Results of the performed numerical experiments are reported for a network with all anti-Hebbian inhibitory neurons {\bf (B)}; with all Hebbian inhibitory neurons {\bf (C)}; with $50$\% anti-Hebbian and $50$\% Hebbian inhibitory neurons {\bf (D)}. 	 
	Raster plots display the firing times of excitatory (red dots) and inhibitory (blue dots) neurons during the simulations. 
	Matrices represent the temporal evolution of the connection weights at different times: $t=0$s (random initialization of the weights), $t=20$s (middle of the learning phase) and $t=40$s (end of the learning phase). The color denotes if the connection is excitatory (red), inhibitory (blue) or absent (white) and the color gradation the strength of the synaptic weight. The final configuration of the connection weights is shown schematically on the right in each case.
	\label{fig:Figure1}
	} 
\end{figure*}


\noindent We begin by investigating the necessary conditions for the emergence of modular structure as induced by training with external stimuli. As a general set-up, we consider heterogenous globally coupled networks composed of $N = 100$ QIF neurons, each neuron receiving weak and independent Gaussian noise. The neurons with labels in the interval $i \in [0:79]$ are excitatory and those labelled $i \in [80:99]$ are inhibitory. We consider the neurons to be non-identical with their excitabilities normally distributed, leading to spontaneous firing frequencies in the range $[0, 8]$ Hz. Synaptic weights $w_{ij}$ from the $j$-th pre-synaptic neuron to the $i$-th post-synaptic one are subject to STDP, with weights bounded in the interval $w_{ij} \in [0,1]$ for excitatory pre-synaptic neurons and in the interval $w_{ij} \in [-1,0]$ for inhibitory ones. See Methods for more details. 

The stimulation protocol consists of three stages, as illustrated in Fig.~\ref{fig:Figure1}A : an initial resting phase followed by a learning period and a final consolidation phase. The simulations are performed by ensuring the model respects some biologically realistic conditions: ($i$) the networks are allowed to continue their spontaneous activity after the learning phase; ($ii$) the adaptation of the synaptic weights is always active throughout the whole simulation i.e., before, during and after the learning.

Initially, despite the network being globally coupled the weights are randomly assigned with small positive (negative) values for excitatory (inhibitory) synapses. The system is left to relax for five seconds in the absence of external stimuli. During this stage the network stabilizes into an asynchronous state with the neurons firing irregularly at low frequencies, as shown in the raster plots corresponding to the time interval $[0:5]$ seconds in Fig.~\ref{fig:Figure1}. The emergence of the asynchronous irregular dynamics in our globally coupled model is due to the heterogenous distribution of the excitabilities and of the weights, and  the external noise. However, a similar dynamics can be observed in networks of identical neurons (same excitabilities) with random connectivity, as shown in Supplementary Sec.~\ref{sec:supp6} and in Refs.~\cite{brunel2000dynamics,litwin2014,yang2024}.

During the learning phase two independent stimuli are applied, each targeting a different (non-overlapping) neuronal population, as shown in Fig.~\ref{fig:Figure1}A. This is done in order to mimic the segregation of (sensory) information being projected to nearby but separate neuronal populations, and to study the role of this segregation for the emergence of modular neuronal architectures. In particular, one stimulus targets population P$_1$ consisting of the first half of excitatory and inhibitory neurons labelled as E$_1 := \{i \in [0:39] \}$ and $I_1 := \{i \in [80:89]\}$. The second stimulus targets population P$_2$ consisting of the second half of excitatory and inhibitory neurons, $E_2 := \{i \in  [40:79] \}$ and $I_2 := \{i \in [90:99]\}$. 

The two stimuli are applied for a time duration of one second, randomly alternating between the two populations P$_1$ and P$_2$. During each stimulation period, a constant external positive current is applied to the selected population for $800$ ms. The stimulus triggers the target neurons to fire at about $50$ Hz. After $800$ ms the external current is turned off and the network is left to relax for other $200$ ms in order to prevent temporal correlations when alternating between stimulated populations. This protocol is repeated $35$ times for a total of 35 seconds. Once the training phase is finished, the network is allowed to evolve freely for $20$ seconds in the absence of stimuli. During this phase, however, the synaptic adaptation remains active thus affecting the stabilization of the learned patterns.

\subsection*{Role of Hebbian and anti-Hebbian learning for the emergence of modular connectivity}

\noindent To explore the role played by the inhibitory neurons in the learning process, we study three different scenarios: ($a$) all inhibitory neurons follow anti-Hebbian STDP, ($b$) all inhibitory neurons follow Hebbian learning, and ($c$) a mixed situation where 50\% are Hebbian and 50\% anti-Hebbian. 
 
\noindent \emph{(a) Anti-Hebbian inhibition.} 
In the case with only anti-Hebbian inhibitory plasticity, the network develops into a \emph{winner-takes-all} architecture promoting the competition of the two sub-populations P$_1$ or P$_2$ on the other one, as shown in Fig.~\ref{fig:Figure1}B. 
In order to describe the mechanism leading to this architecture, imagine that population P$_1$ is stimulated, consequently all its neurons become active and fire with high frequency. Therefore all the connections E$_1 \to \{\textrm{E}_1,\textrm{I}_1\}$ are reinforced due to the Hebbian nature of the pre-synaptic excitatory neurons. At the same time, all the inhibitory synapses I$_1 \to \{\textrm{E}_1,\textrm{I}_1\}$ are weakened because they are anti-Hebbian. During the stimulation of P$_1$, the neurons in populations P$_1$ and P$_2$ are far from being mutually synchronized, therefore the connections E$_1 \to \{\textrm{E}_2,\textrm{I}_2\}$ are weakened while I$_1 \to \{\textrm{E}_2,\textrm{I}_2\}$ reinforce. The random alternation of the stimuli to populations P$_1$ and P$_2$ induces a gradual emergence of a modular structure, as visible in the connectivity (weight) matrices at times $t = 20$ sec and $t = 40$ sec in Fig.~\ref{fig:Figure1}B. 

The resulting architecture promotes the competition between the two excitatory populations E$_1$ and E$_2$ and the alternating prevalence of one of them. For example, stimulation to E$_2$ would activate the companion inhibitory neurons through the \emph{feed-forward} connections ($\textrm{E}_2 \to \textrm{I}_2$) which, in turn, would shut down all neurons in P$_1$ via the strong I$_2 \to \{\textrm{E}_1,\textrm{I}_1\}$ connections. This is visible in the raster plot in Fig.~\ref{fig:Figure1}B : as the training progresses ($t = 5 - 40$ sec), the neurons of the stimulated population fire at high frequency but the neurons of the non-stimulated one are silenced. 

\noindent \emph{(b) Hebbian inhibition.} 
In the presence of only Hebbian inhibitory neurons the network also develops a modular organization, however in this case the two populations become disconnected from each other, Fig.~\ref{fig:Figure1}C. Analogously to the previous case, the stimulation of one population (e.g., P$_1$) results in the strengthening of the intra excitatory connections (increase of $\textrm{E}_1 \to \{\textrm{E}_1,\textrm{I}_1\}$ synaptic weights) and the weakening of the inter excitatory connections across different populations (decrease of $\textrm{E}_1 \to \{\textrm{E}_2,\textrm{I}_2\}$ weights). However, the Hebbian inhibitory plasticity now induces the strengthening of the internal inhibitory connections $\textrm{I}_1 \to \{\textrm{E}_1,\textrm{I}_1\}$ and the weakening of the inhibitory synapses across populations. The disconnection of the two populations happens gradually during the learning phase. As shown in the raster plot in Fig.~\ref{fig:Figure1}C, during the initial stimulation epochs, the stimulated population shuts down the activity of the non-stimulated population. But as the training proceeds, the two populations detach one from the other and the non-stimulated population begins to display a low firing resting-state activity. 

\noindent \emph{(c) Mixed inhibition.} 
In the last case with mixed anti-Hebbian and Hebbian inhibitory neurons, the network also develops two modules, but, as shown in Fig.~\ref{fig:Figure1}D, the resulting connectivity is a combination of the two configurations observed in the previous cases. The Hebbian neurons form self E--I loops within each population (i.e. $\textrm{E}_1 - \textrm{I}_1^H$ and $\textrm{E}_2 - \textrm{I}_2^H$). This internal feedback inhibition avoids that the excitatory neurons fire at too large rates. Meanwhile, the anti-Hebbian inhibitory neurons form lateral, feed-forward connections which shut down the firing of the other population, in other terms the sub-population $\textrm{I}_1^A$ ($\textrm{I}_2^A$) inhibits all neurons in P$_2$ (P$_1$).

\subsection*{Resting-state network dynamics after learning}

So far, we have shown that selective stimulation to distinct populations consistently gives rise to modular networks and that the resulting configuration depends on the type of plasticity affecting the inhibitory neurons. However, the relevance or plausibility of a learning model should also require that the network exhibits a biologically meaningful dynamical behaviour after training~\cite{del2003,amit2003,litwin2014,zenke2015,ocker2019,fauth2019,yang2024}.

\noindent \emph{(a) Anti-Hebbian inhibition.} In the simulations with anti-Hebbian inhibitory STDP, the neuronal activity in the post-learning stage ($t > 40$ sec) is dominated by one of the two populations. In Fig.~\ref{fig:Figure1}B,  P$_2$ is active and P$_1$ is silent. But this randomly changes over realizations. In general, once the training stage is finished, all neurons start to fire spontaneously driven by the background Gaussian noise. The excitatory sub-population that attains a larger level of internal activity earlier is the one that wins. 
Due to the lack of internal feedback inhibition ($\textrm{I}_1 \nrightarrow \textrm{E}_1$ or $\textrm{I}_2 \nrightarrow \textrm{E}_2$) the winning excitatory sub-population fires rapidly and the companion inhibitory neurons (strongly stimulated) suppress the activity of the other population: e.g. if the neurons in E$_2$ fire faster than those in E$_1$ the sub-population I$_2$ inhibits both E$_1$ and I$_1$. 

\noindent \emph{(b) Hebbian inhibition.} In the case with Hebbian inhibitory neurons, we found that the two populations P$_1$ and P$_2$ become independent, each forming a \emph{feedback E--I loop}. The presence of the internal feedback inhibition allows the two populations to settle into a resting-state behaviour after training that is characterized by a low firing frequency of approximately $1.0$ Hz, as visible in the raster plot of Fig~\ref{fig:Figure1}C, for $t > 40$ sec. Interestingly, brief events of internal partial synchrony of the two populations are also observed. While in the case with only anti-Hebbian inhibition such a synchronized event would trigger the excitatory neurons to permanently increase their firing, here the presence of the internal feedback inhibition ($\textrm{I}_1 \to \textrm{E}_1$ and $\textrm{I}_2 \to \textrm{E}_2$) avoids the constant synchronization of excitatory neurons, while it keeps their activity at low frequency. 

\noindent \emph{(c) Mixed inhibition.} The post-learning behaviour in the mixed scenario is very similar to the Hebbian case, as shown in Fig.~\ref{fig:Figure1}D. However, in this case populations P$_1$ and P$_2$ are not independent but they inhibit each other. As a consequence, an event of partial synchrony occurring in one of the two modules temporarily shuts down the other module, avoiding their mutual synchrony. Given that the plasticity remains active during this post-learning spontaneous activity, the connection weights continue to be updated. As we will discuss in the following, the occurrence of these spontaneous synchronization events plays a relevant role in the maintenance of the learned memories. 

In summary, despite modular organization of the connectivity was found to emerge in the three considered scenarios, only the combination of anti-Hebbian and Hebbian inhibitory neurons resulted in a network satisfying all the desired biologically plausible properties. Anti-Hebbian inhibition alone led to a network with unrealistic post-learning dynamics. Hebbian inhibition alone gives rise to a biologically meaningful resting-state behaviour but at the cost of splitting the network into two disconnected populations. Therefore, in the following we will limit to consider the mixed scenario, with both anti-Hebbian and Hebbian inhibition.

The numerical experiments presented so far were repeated for a variety of special cases in order to validate the robustness of the results and the model: namely, (i) some of the neurons are not trained by stimuli (see Supplementary Sec.~\ref{sec:supp1}), (ii) at every stimulation epoch random subsets of neurons are targeted (Supplementary Sec.~\ref{sec:supp2}), (iii) the intensity of the stimulation current is randomly fluctuating (Supplementary Sec.~\ref{sec:supp3}), (iv) the network is composed of a large number of neurons, and (v) the neurons are randomly connected with an Erd\"os-R\'enyi distribution of the directed links (Supplementary Sec.~\ref{sec:supp6}).

\section*{Spontaneous recalls support consolidation and long-term maintenance of memories}
 
 \begin{figure*}[p]
	\centering
	\includegraphics[width=0.85\textwidth]{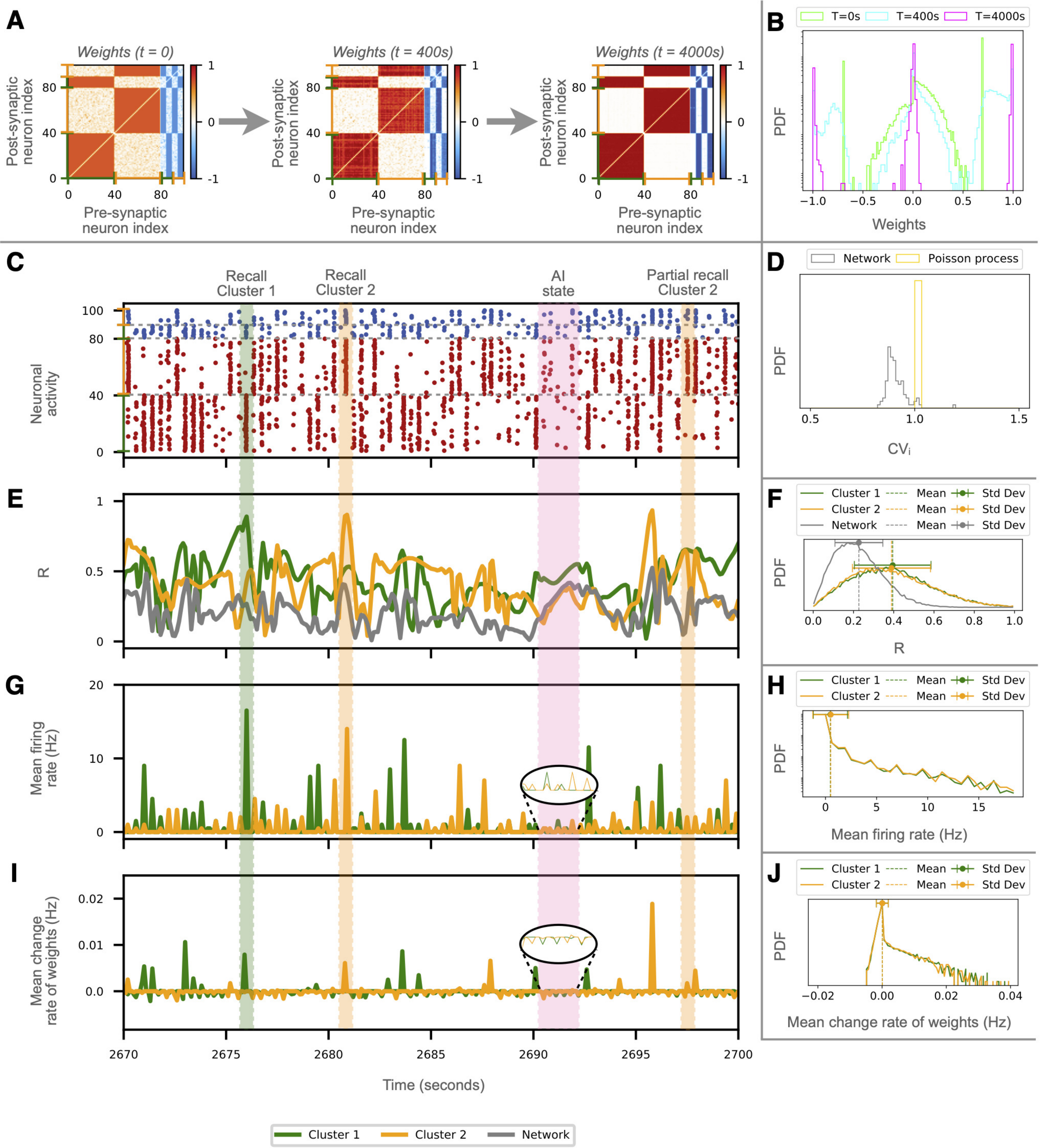}
	\caption{
	{\bf Consolidation of imperfectly learned memories.}
	{\bf (A)} Temporal evolution of the connectivity matrix during spontaneous activity in the absence of stimulation. Initial connectivity with two unfinished modules at $t=0$ is reinforced over time. The excitatory (inhibitory) connections are marked as in Fig.~\ref{fig:Figure1}.
	{\bf (B)} Evolution of the distribution of link weights (probability density functions, PDFs, in a linear-logarithmic scale) in the connectivity matrices at $t=0$s (light green), $t=400$s (cyan) and $t=4000$s (magenta).
	{\bf (C-J)} Evolution of the network activity and various metrics in absence of stimulation for a sample of 30 seconds. Some spontaneous recalls are highlighted, for population P$_1$ (green shade) and P$_2$ (orange shade). Pink shadow marks an epoch of asynchronous irregular firing activity without recalls. 
	{\bf (C)} Raster plot with excitatory (inhibitory) neurons marked in red (blue).
	{\bf (D)} PDFs of the coefficient of variations $CV_i$ for all the neurons during the entire simulation (grey) and for a homogeneous Poisson process (yellow).
	{\bf (E)} Instantaneous Kuramoto order parameters $R$ for the networks (gray) and for neurons in population P$_1$ (green) and P$_2$ (orange), and their corresponding PDFs over the entire simulation {\bf (F)}.
	{\bf (G)} Temporal evolution of the mean firing rates for populations P$_1$ and P$_2$, and {\bf (H)} their corresponding frequency distributions (in linear-logarithmic scale) showing a peak at $2$ Hz and long time tail.
	{\bf (I)} Instantaneous change rates of synaptic weights in both populations P$_1$ and P$_2$ over time and their PDF (in linear-logarithmic scale) {\bf (J)} over the entire simulation time, showing the prevalence of positive weight changes (reinforcement) with respect to negative ones (depression).
	\label{fig:Figure2}	
	} 
\end{figure*}

Given that in our model plasticity is not frozen after training---but it is left active afterwards---we now investigate the potential role of the spontaneous events observed during the post-learning resting-state for the consolidation and maintenance of the memories. We will refer to these events as {\it memory recalls}. First, we will show that the recalls facilitate the completion of imperfectly learned memories and, second, we will study their role for the regeneration of memories that have been partially lost.

\subsection*{Memory consolidation}  

In order to mimic a hypothetical scenario in which the training stage would stop before completion, we prepared an initial weight matrix representing an imperfectly learned connectivity structure. An example of this weight matrix is shown at $t=0$ in Fig.~\ref{fig:Figure2}A. Specifically, the intra-modular synaptic weights are set to $w_E = 0.7$ for excitatory and to $w_I = -0.7$ for inhibitory connections. The inter-modular excitatory (inhibitory) connections are chosen randomly from a Gaussian distribution with mean $0$ and standard deviation $0.15$ restricted to positive (negative) values.

Once fixed the initial weight matrix, the activity of the network is then left evolve spontaneously, driven only by the background Gaussian noise. However, we observe that the modular organization of the network is reinforced: the intra-modular synaptic weights of the two populations are strengthened and the inter-modular synapses are weakened, as shown by the weight matrices displayed in Fig.~\ref{fig:Figure2}A and the corresponding weight distributions in Fig.~\ref{fig:Figure2}B at $t=400$ sec and $t=4000$ sec.

The mechanism allowing for the completion of the connectivity pattern is summarized in Figs.~\ref{fig:Figure2}C-J. Panel C shows a 30 seconds sample of activity of the network. This is characterized by an asynchronous irregular evolution with low firing spontaneous activity joined to occasional spontaneous recalls occurring at random times. This observation is confirmed by the fact that the background activity is characterized by a distribution of the coefficient of variations CV$_i$ of the neurons in the interval $[0.8:1.0]$ (panel D, grey distribution) lying close to a Poisson process (in yellow), thus resembling the irregular activity observed in the cortex {\em in vivo}~\cite{perkel1967neuronal}. The synchronization order parameter of the network $R$ fluctuates in time with values around $0.2$ (panel F, gray distribution), meaning that the network is far from being synchronized. As a matter of fact, for an asynchronous network of $N=100$ neurons, due to finite size fluctuations the expected value of the order parameter will be not far from the one here measured, namely $R \simeq 1/\sqrt{N} = 0.1$. Finally, the distribution of firing rates displays a main peak around $2$ Hz and an exponential tail reaching at most $20$ Hz (panels G and H).

During spontaneous recalls, the transient increase in synchrony and firing rates of a subset of the neurons (in one of the two clusters) triggers the activation of their synaptic adaptation (panel I), reshaping the synaptic weights and completing the formation of the modular patterns. The specific neurons participating in a recall changes from event to event but they tend to involve a majority of neurons of either population P$_1$ or P$_2$; see for example the events highlighted by green and orange shadows. The recalls coincide with transient peaks of the order parameter and of the mean firing rates of the populations involved (panels E and G, green lines). Consequently, P$_1$ and P$_2$ are internally more synchronized than the network average, with their order parameters fluctuating around $0.4$ (panel F), and a broad distribution of firing rates (panel H). The similarity between the PDFs for populations P$_1$ and P$_2$ evidences that recalls occur---on average---with the same probability in both populations. Over time, the net effect of recalls is that the intra-modular synapses between neurons of the same population become reinforced while the inter-modular connections are weakened.

At this point, we remind that our model incorporates a natural, slow tendency to forget the value of the synaptic weights. Therefore, the question arises of how often should spontaneous recalls happen---and how strong should they be---in order to avoid the forgetting of the learned memories. In the example considered here, the weights between the neurons of a population undergo a natural depression during epochs of the resting-activity without recalls (pink shadow in Fig.~\ref{fig:Figure2}C). 
For the used parameters, a forgetting term of $f = 0.1$ and an average spontaneous firing rate of $2$ Hz, we estimate that a period of $\simeq 11$ seconds of asynchronous firing would be required to forget the contribution of a single recall event (for more details on the calculus see Supplementary Sec.~\ref{sec:supp9}). The dominance of the reinforcement during recalls with respect to depression during the asynchronous irregular epochs is confirmed by the asymmetry in the distribution of positive weight change rates over negative ones, Fig.~\ref{fig:Figure2}J.
 
In summary, we can conclude that, in our model, spontaneous recalls happening during resting-state activity allow for the consolidation of imperfectly learned memories as well as for their long-term maintenance against natural forgetting~\cite{litwin2014,yang2024}.

\subsection*{Regeneration of damaged excitatory and inhibitory connectivity} 

\begin{figure*}[ht!]
	\centering
	\includegraphics[width=0.90\linewidth]{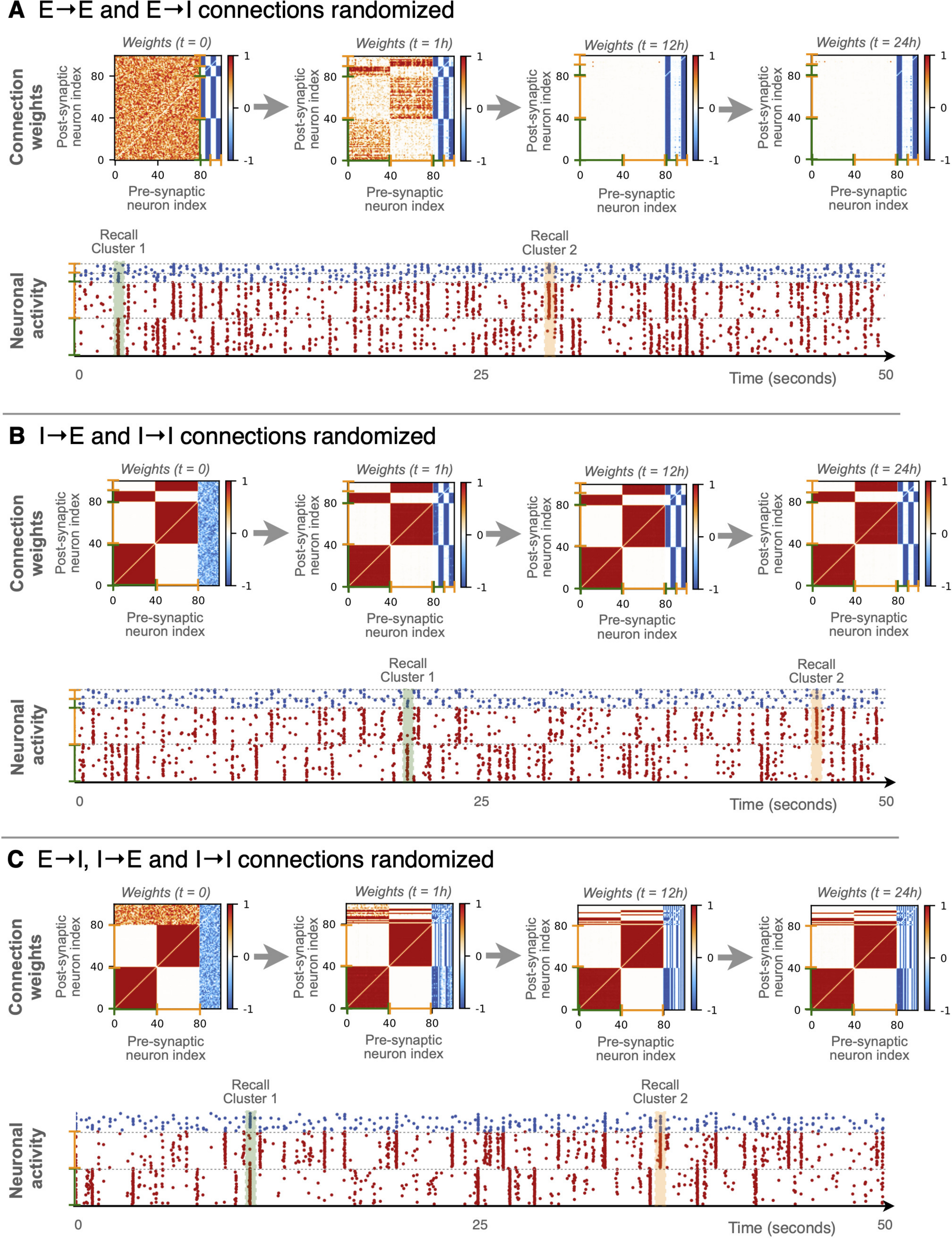}
	\caption{
	{\bf Recovery of damaged modular structure by spontaneous recalls.} \\
	{\bf A} Recovery experiment starting from randomized excitatory {\bf (A)}, inhibitory {\bf (B)}, or inhibitory plus excitatory-inhibitory {\bf (C)} synaptic connections. 
	 In all the three cases, we study the partial recovery of the original modular organization of the synaptic weights over time. This recovery is mediated in all the cases by the emergence of spontaneous recalls, ---transient events of partial synchronization between neurons associated with one of the two originally stored memories, highlighted in green for cluster 1 and orange for cluster 2.
	\label{fig:Figure3}
	} 
\end{figure*} 

Empirical observations have shown that excitatory synapses are more volatile than inhibitory ones~\cite{defelipe1997inhibitory,Rubinski2015remodeling}, leading to the hypothesis that reorganization of excitatory connections might be associated with short-term plasticity while inhibitory adaptation might support the long-term maintenance of those memories~\cite{karunakaran2016PVplasticity,barron2017inhibitory,mongillo2018inhibitory,giorgi2021roles}. We now test the plausibility of this idea in our model. 

We consider a first case where the weights of the excitatory synapses ($w_{EE}$ and $w_{IE}$) are initially randomized between 0 and 1, but the values of the weights of the inhibitory synapses coincide with those obtained as after the training stage. See the weight matrix for $t=0$ in Fig.~\ref{fig:Figure3}A. The network is then let to evolve spontaneously for $24$ hours, driven only by the background Gaussian noise. After one hour of simulated time, the excitatory connections have partially recovered the original modular pattern (see the weight matrix for $t = 1h$ in Fig.~\ref{fig:Figure3}A). This recovery is clearly mediated by the spontaneous recalls as illustrated in the raster plot in Fig.~\ref{fig:Figure3}A. Whenever a sufficiently large group of excitatory neurons (associated to one of the previously learned memories) fire in a short time window, this event activates the inhibitory neurons associated to the corresponding memory items and triggers a partial recall via the feedback and feed-forward mechanism previously explained. Consequently, the excitatory synapses involved in the recalled memory are partially reinforced while all other connections are weakened. The structure of the inhibitory weights appears to be deteriorating due to the more disordered activity observed in the network. As a result, in the long-term ($t=12h$ and $t=24h$), the modular connectivity structure encoding the stimuli is completely lost. Therefore, the maintenance of the inhibitory weight structure is sufficient for the (partial) recovery of the memory items over an intermediate time scale only.

The experiment was repeated but randomizing only the inhibitory weights ($w_{EI}$ and $w_{II}$, the corresponding weight matrix is shown at $t=0$ in Fig.~\ref{fig:Figure3}B), the memorized pattern is fully recovered and maintained at all times. The main reason for this difference is related to the fact that when the excitatory synapses are preserved, the network displays a more sparse activity than with initially randomized excitatory connections. In the latter case, there are more interactions among excitatory neurons and this gives rise to a background activity involving most of the excitatory neurons. Therefore, one finally obtains a weight matrix with a more disordered structure (see in Fig.~\ref{fig:Figure3}A weight matrix at $t=1h$), which becomes unstable after several hours of spontaneous activity. Another reason of this difference is related to the asymmetric 80-20\% $E/I$ ratio, leading to notably different spontaneous firing frequencies---and recall probabilities---when either the excitatory or the inhibitory synapses are randomized.

Finally, the experiment is repeated by randomizing all weights except the excitatory to excitatory connections ($w_{EE}$), in Fig.~\ref{fig:Figure3}C. In the initial moments of the simulation, again we observe the spontaneous recall of excitatory neurons, but the inhibitory neurons due to the random distribution of the $E \to I$ connections are now randomly associated to one of the two clusters. This leads to a slightly different organization in the connectivity matrix than before, but with a similar architecture.

These results show that the conservation of the excitatory to excitatory $E \to E$ connections is indeed sufficient to maintain and reconsolidate the structure. Furthermore, our analysis suggests that memory items can be encoded in different types of connections with different effects on the intermediate or long-term storage and on the efficiency of recovering, which may echo the different types of memories present in the brain.

\section*{Generalization of the model to more complex situations}

So far, we have illustrated the results for networks trained with two stimuli, applied at separate neuronal sub-populations. We will now show that the model can be generalized. Firstly, the model is trained to an increasing number of memories in order to estimate its memory capacity. Secondly, the stability of the model is validated for larger network sizes and more stimuli, and lastly, we demonstrate the emergence of hub neurons by training the network with overlapping memory items.

\subsection*{Memory capacity of the model}

\begin{figure*}[p]
	\centering
	\includegraphics[width=0.82\textwidth]{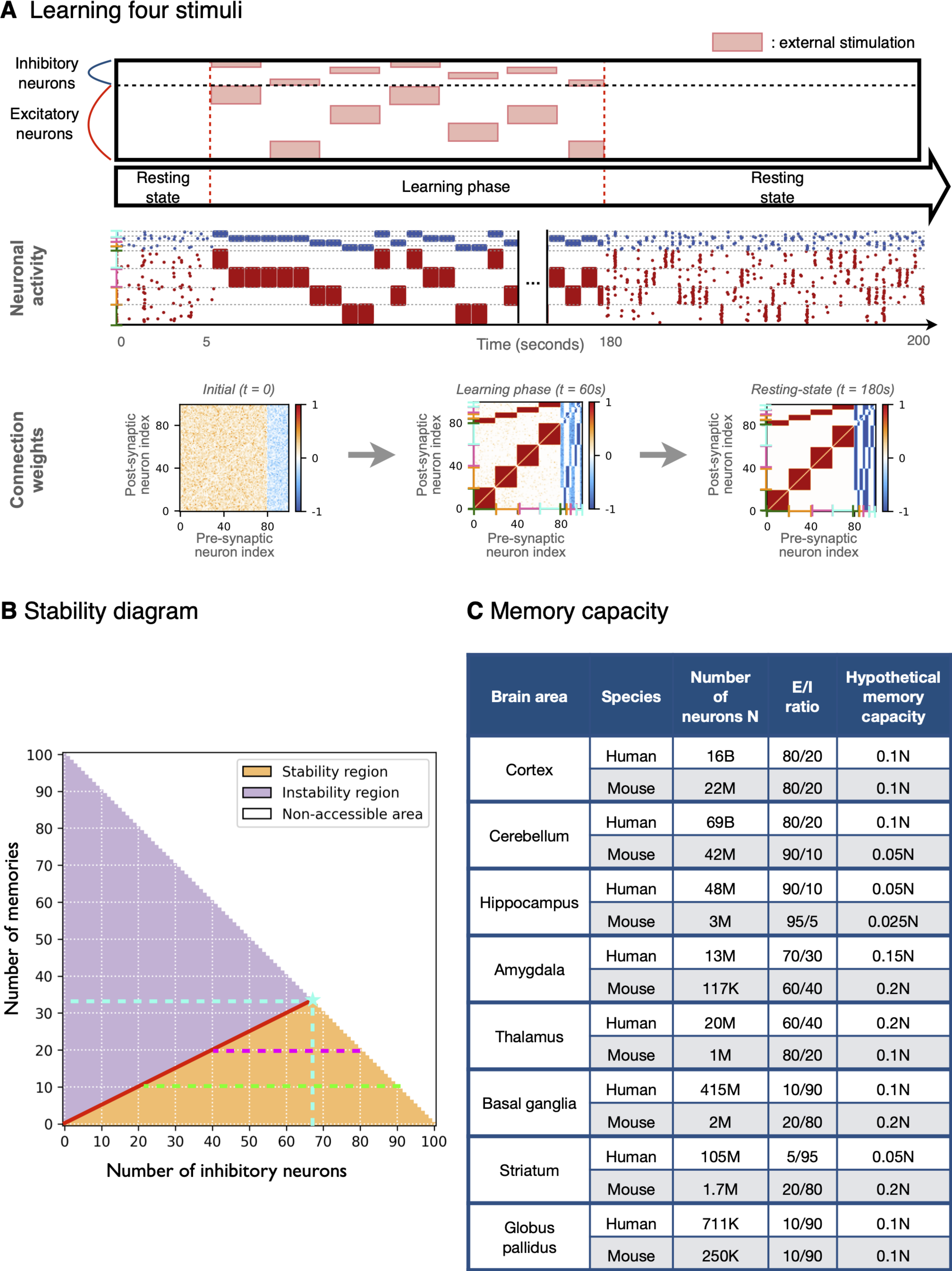}
	\caption{
	{\bf Memory capacity of the model.} 
	{\bf (A)} Training the network with $M=4$ non-overlapping stimuli. The green, orange, pink and cyan brackets represent clusters 1, 2, 3 and 4. Results are qualitatively the same as for the example with two memory items.
	{\bf (B)} Stability diagram of the model for a network of $N=100$ neurons. The red line $M = N_I/2$ separating the stability (orange) and instability (purple) regions, represents an upper limit for the number of inhibitory neurons ($N_I=2M$) needed to maintain $M$ independent memory items. The number of excitatory neurons in each realization is $N_E = 100 - N_I$ delimiting the non-accessible areas (white region) of the diagram. The star symbol marks the maximal memory capacity of the network, corresponding to $M^{\star}=33$ items with $N_I^\star = 66$ inhibitory neurons (cyan lines, 33 anti-Hebbian and 33 Hebbian). Magenta and light-green lines highlight the ranges of number of inhibitory neurons (and therefore of the $N_I/N_E=N_I/(100-N_I)$ ratios) that allow the stabilization of $M=20$ and $M=10$ memory items, respectively.
	{\bf (C)} Table comparing the number of neurons and $E/I$ ratios of different parts of the human~\cite{arcelli1997gabaergic,avino2018neuron,d2018physiology,gandolfi2023full,hardman2002comparison,herrero2002functional,karlsen2011total,pelkey2017hippocampal,sah2003amygdaloid,schroder1975morphometrical,singh2016role,spampanato2011interneurons,von2016search} and mouse~\cite{ero2018cell,rodarie2022method,roussel2023mapping} brain and their hypothetical memory capacity obtained by extrapolating the results of our model.	
	} 
	\label{fig:Figure4}
\end{figure*}

We begin by repeating the protocol in Fig.~\ref{fig:Figure1}D but now training the network to $M=4$ non-overlapping stimuli, as shown in Fig.~\ref{fig:Figure4}A. We observe the same qualitative results as in the case with 2 stimuli, the only difference is that now four modules emerge in the connectivity matrix. Also in the present situation, the four neuronal assemblies display spontaneous recalls during the post-learning resting-state phase. A minor difference is that now the 20 inhibitory neurons split into 8 sub-populations, each made of 2 -- 3 anti-Hebbian and Hebbian neurons. This opens the question of the memory capacity of the network, which seems limited by the number of inhibitory neurons.

In order to estimate the memory capacity of our model, we examine the ability of the network to store and recall an increasing number of memory items while varying the ratio of excitatory to inhibitory neurons. Specifically, connectivity matrices were initialized containing an arbitrary number $M$ of memory items (modules), from $M=0$ to $M=100$, and by varying the number of inhibitory neurons from $N_I=0$ to $N_I=100$. Since the size of the network is maintained constant to $N=100$, the number of excitatory neurons is varied consequently as $N_E = 100 - N_I$. 
A network storing $M$ memory items is considered stable if (i) it displays asynchronous irregular firing and (ii) all the memory modules exhibit independent spontaneous recalls. A network with $M$ memories is considered unstable if at least one of the modules does not exhibit recalls and therefore disappears in the long-term. 

The results are summarized in the stability diagram in Fig.~\ref{fig:Figure4}B. The red line separating the stable and the unstable regimes evidences that at least $2M$ inhibitory neurons are needed to hold $M$ memories. This result implies that the network can learn and stabilize a maximum of $M^{\star} = N \,/\, 3 =33$ memories (marked by a cyan star in Fig.~\ref{fig:Figure4}B). In particular, this maximal capacity corresponds to the case with $N_I = 66$ inhibitory neurons, 33 of them anti-Hebbian and 33 Hebbian, and $N_E=33$ excitatory neurons. A simulation for this limit case is shown in Supplementary Sec.~\ref{sec:supp7}. Therefore, we conclude that in our model, the minimal condition for a memory to be robustly encoded and recalled is that the memory is associated to a triplet, made of at least one excitatory neuron, one Hebbian and one anti-Hebbian inhibitory neuron. For example, to store $M=20$ memories (magenta dashed line in Fig.~\ref{fig:Figure4}B) the network needs to contain at least $20$ excitatory neurons and $40$ inhibitory neurons (20 Hebbian and 20 anti-Hebbian). 

In sub-optimal scenarios, we find that for a given number of memory items $M < M^{\star}$, different $E / I$ ratios can guarantee the long-term maintenance of the memories (connectivity patterns). All the possible $E \,/\,I$ ratios respecting these constraints can give rise to configurations in which the $M=20$ memories are robustly stored. This amounts to have the excitatory to inhibitory ratio ranging from $ N_E \,/\, N_I = 20 \,/\, 80 = 0.25$ up to $60 \,/\, 40 = 1.5$, where the total number of neurons is maintained equal to $N=100$. For a case with $M=10$ memories (green dashed line), the interval of acceptable ratios widens from $N_E \,/\, N_I = 10 \,/\, 90 = 0.111$ to $80 \,/\, 20 = 4$.  
In summary, given that each memory requires a triplet of neurons, the memory capacity of our model is limited by the size of the smaller population of neurons, i.e. $\min\left[ N_E, N_I^A, N_I^H \right]$. For a network with the typical E to I ratio of $80 / 20$, it implies that its memory capacity is controlled by the number of inhibitory neurons instead of the number of excitatory neurons.

The analysis of a few test cases reported in Supplementary Sec.~\ref{sec:supp4} for $M=4$ memory items confirms the validity of the upper limit $M = 2 N_I$. In particular, in Supplementary Sec.~\ref{sec:supp4} we studied the following situation: (i) each memory pattern is associated to one Hebbian and one anti-Hebbian inhibitory neuron, (ii) one memory pattern is missing an anti-Hebbian inhibitory neuron, and (iii) one memory item is missing one Hebbian inhibitory neuron.

The flexible relation between memory capacity and $E/I$ ratios found in our model invites to question why would nature favour the specific $E/I$ ratios observed in the mammalian brains, and whether these ratios might be related to their memory storage. To explore this question we computed the hypothetical memory capacities of the human and mouse brains, extrapolating the predictions from our model. Moreover, given that different parts of the brain are made of distinct number of neurons and $E/I$ ratios, we extended the calculations to distinguish the cortex, the cerebellum and a few subcortical regions. The results are shown in Fig.~\ref{fig:Figure4}C.
The $80/20$ excitatory to inhibitory ratio---which is mostly representative for the cerebral and cerebellar cortices---implies a hypothetical memory capacity of $M=0.1 N$, well below the theoretical maximal capacity of $M^{\star}=0.33N$. Still, given that the cortex and the cerebellum contain most of the neurons in the brain, in absolute terms this would imply that the human cortex and cerebellum have memory capacities of $M^{ctx} \approx 1.6 \times 10^9$ and $M^{cer} \approx 7.0 \times 10^9$ respectively, and $M^{ctx} \approx 2.2 \times 10^6$ and $M^{cer} \approx 10.5 \times 10^6$ in mice.
Rather surprisingly, we find the lowest memory capacity in the Hippocampus, with $M = 0.050N$ and $M = 0.025N$ in humans and mice due to a reduced fraction of inhibitory neurons. The amygdala lies among the regions with largest memory capacity, $M=0.15N$ and $M=0.2N$ for humans and mice. This particularity of the amygdala could be related to the fact that this area is involved in the formation of emotional memories~\cite{mcgaugh2004amygdala}. Nevertheless, the amygdala contains few neurons as compared to other regions, hence its absolute capacity would be definitely smaller.

The stability of a neuronal network can be compromised by the network size and the number of memories stored. Hence, to validate the robustness of our model we considered a network of 1000 neurons trained with 10 stimuli, Fig.~\ref{fig:Figure5}A. After training, the network was left to evolve spontaneously for a period of 4 hours. 
In Fig.~\ref{fig:Figure5}B, we observe the formation of 10 modules among the excitatory connections, each one associated with one group of Hebbian and one group of anti-Hebbian inhibitory neurons. The weight matrices remain practically unaltered over time up to 4 hours. This is additionally evidenced by the stability of the mean inter- and intra-cluster weights, see Fig.~\ref{fig:Figure5}C. 

The stability of the learned patterns depends on a few ingredients. On the one hand, the modules remain stable without merging or drifting~\cite{manz2023}. This is possible because the anti-Hebbian inhibitory neurons ensure that recalls only occur one module at a time, depressing the other modules. If two modules were to undergo a recall simultaneously, their mutual synapses would become reinforced, causing their merging over repeated simultaneous events. Indeed, the spike counts remain below 5\% of the network size, meaning that more memories are never fully recalled at the same time and that the network is far from being fully synchronized. This despite the occurrence of sporadic peaks of activity spread over the whole network, which however appear to be quite sparse (see the zoom of the grey shadowed interval in Fig.~\ref{fig:Figure5}A). A closer look at various recall events (see the enlargements of the coloured intervals in Fig.~\ref{fig:Figure5}A) shows that, usually, only the neurons of a given memory fire within a time window of $0.2$ seconds.

On the other hand, for the modules to be maintained, a sufficient frequency of recalls is necessary. As just mentioned, only one module at a time can display a recall. Thus, the more memories are stored, the less frequently each module will be recalled. In consequence, the long-term maintenance of the modules is possible as long as the forgetting term $f$ is sufficiently small, and inversely proportional to the number of modules $M$, (see Supplementary Sec.~\ref{sec:supp9} for a more detailed discussion).
This experiment has also been repeated for different network sizes and number of stimuli, see Supplementary Sec.~\ref{sec:supp8}.

\subsection*{Robustness of the model for larger network sizes trained with multiple stimuli}

\begin{figure*}[ht!]
	\centering
	\includegraphics[width=0.85\linewidth]{Figure5.pdf}
	\caption{
	{\bf Stability of a network of $N=1000$ neurons trained with $M=10$ stimuli.} 
	{\bf (A)} Post-learning raster plots of the neuronal activity at times: t=0h, t=2h and t=4h. In each plot, the spike count, normalized to the total number of neurons $N$ and estimated over time bins of $0.2$ sec, is displayed below the time axis. Selected spontaneous recalls are highlighted by colored shadows while a peak of global activity is highlighted by a grey shadow. Some of these events are zoomed on the right side of the panels.
 	{\bf (B)} Post-learning weights matrices at times: t=0h, t=2h and t=4h.
 	{\bf (C)} Post-learning evolution of mean intra- (solid lines) and inter- (dashed line) clusters weights for excitatory to excitatory (E-E dark red), excitatory to inhibitory (E-I red), anti-Hebbian inhibitory to excitatory (A-E dark blue), anti-Hebbian inhibitory to inhibitory (A-I cyan), Hebbian inhibitory to excitatory (H-E blue) and Hebbian inhibitory to inhibitory (H-I steel blue) connections.	} 
	\label{fig:Figure5}
\end{figure*}

The stability of a neuronal network can be compromised by the network size and the number of memories stored. Hence, to validate the robustness of our model we considered a network of 1000 neurons trained with 10 stimuli, Fig.~\ref{fig:Figure5}A. After training, the network was left to evolve spontaneously for a period of 4 hours. 
In Fig.~\ref{fig:Figure5}B, we observe the formation of 10 modules among the excitatory connections, each one associated with one group of Hebbian and one group of anti-Hebbian inhibitory neurons. The weight matrices remain practically unaltered over time up to 4 hours. This is additionally evidenced by the stability of the mean inter- and intra-cluster weights, see Fig.~\ref{fig:Figure5}C. 

The stability of the learned patterns depends on a few ingredients. On the one hand, the modules remain stable without merging or drifting~\cite{manz2023}. This is possible because the anti-Hebbian inhibitory neurons ensure that recalls only occur one module at a time, depressing the other modules. If two modules were to undergo a recall simultaneously, their mutual synapses would become reinforced, causing their merging over repeated simultaneous events. Indeed, the spike counts remain below 5\% of the network size, meaning that more memories are never fully recalled at the same time and that the network is far from being fully synchronized. This despite the occurrence of sporadic peaks of activity spread over the whole network, which however appear to be quite sparse (see the zoom of the grey shadowed interval in Fig.~\ref{fig:Figure5}A). A closer look at various recall events (see the enlargements of the coloured intervals in Fig.~\ref{fig:Figure5}A) shows that, usually, only the neurons of a given memory fire within a time window of $0.2$ seconds.

On the other hand, for the modules to be maintained, a sufficient frequency of recalls is necessary. As just mentioned, only one module at a time can display a recall. Thus, the more memories are stored, the less frequently each module will be recalled. In consequence, the long-term maintenance of the modules is possible as long as the forgetting term $f$ is sufficiently small, and inversely proportional to the number of modules $M$, (see Supplementary Sec.~\\ref{sec:supp9} for a more detailed discussion).
This experiment has also been repeated for different network sizes and number of stimuli, see Supplementary Sec.~\ref{sec:supp8}.

\subsection*{Formation of hubs by training to overlapping memories}

\begin{figure*}[ht!]
	\centering
	\includegraphics[width=0.90\linewidth]{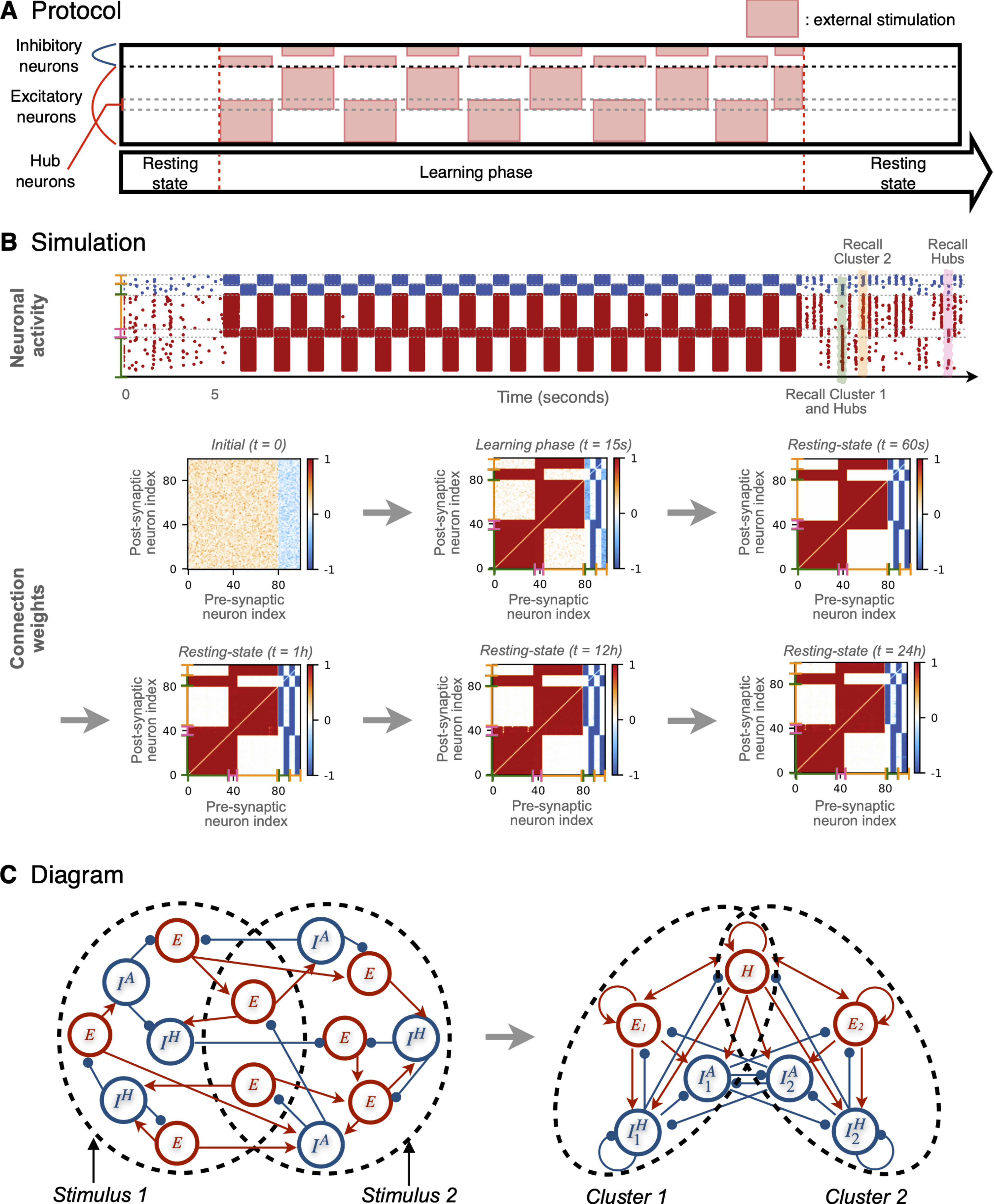}
	\caption{
	{\bf Training the network with overlapping stimuli.} 
	{\bf (A)} Experimental protocol for a network of $N=100$ neurons trained with $M=2$ stimuli that share 8 excitatory neurons. 
	{\bf (B)} Simulation and learning results. Connectivity matrices show the evolution of the synaptic weights leading to the emergence of two modules which overlap over 8 hub neurons. The raster plot shows the simulated neuronal activity of the network during the initial resting phase, the learning stage and the post-learning period. The activity during the post-learning phase is characterized by a variety of spontaneous recall events involving P$_1$ neurons and the hubs (green shadow), P$_2$ neurons without the hubs (orange shadow) and the hubs alone (pink shadow). 
	{\bf (C)} Schematic diagrams representing the formation of the synaptic connectivity with two stable populations of neurons and overlapping hubs. Each population is composed at least of a population of excitatory neurons E$_x$ and a population of hub neurons $H$ (in red), one population of Hebbian inhibitory neurons I$_x^H$ and one population of anti-Hebbian inhibitory neurons I$_x^A$ (in blue) ($x=1,2$).
	Dashed circles identify groups of neurons admitting synchronization events (memory recalls).
	} 
	\label{fig:Figure6}
\end{figure*}

We finish the investigation of the model by considering the case in which two stimuli entrain an overlapping set of excitatory neurons and explore the possibility that these could encode for more than one stimulus~\cite{fauth2019,bergoin2023inhibitory,bergoin2023phd,manz2023,yang2024}. 
This would correspond to a ``mixed selectivity'' scenario as it is often found in neurons of the prefrontal cortex~\cite{rigotti2013importance,parthasarathy2017mixed}. During the training stage, populations P$_1$ and P$_2$ are now allowed to share eight excitatory neurons, see Fig.~\ref{fig:Figure6}A. Furthermore, stimulation to P$_1$ and P$_2$ is strictly alternated in order to facilitate the formation of the connections, instead of selecting randomly between P$_1$ and P$_2$, as done in the previous simulations.

The results in Fig.~\ref{fig:Figure6}B are similar to the ones before only that besides the formation of two clusters, now a set of structural \emph{hub neurons} emerge in the connectivity matrix. These excitatory hubs are strongly connected to both modules. They are weakly affected by the feed-forward anti-Hebbian inhibition (since they belong to both populations), but are notably affected by the feedback Hebbian inhibition of the two populations. Figure~\ref{fig:Figure6}C shows a schematic diagram of the initial and of the resulting connectivity. 

Regarding the post-learning resting phase, see raster plot of Fig.~\ref{fig:Figure6}B, the network displays again a stable low-frequency asynchronous firing activity but with richer spatio-temporal patterns than in the previous examples. The spontaneous recalls are more varied now with events displaying: (i) synchronous spiking of one population including the hubs (event shaded in green), (ii) synchronous spiking of the population without the hubs (event shaded in orange), and (iii) synchronous spiking of the hub neurons alone (event shaded in pink). Supplementary Sec.~\ref{sec:supp5} shows similar results for a network trained with $M=4$ overlapping stimuli. 

In conclusion, the introduced model can be generalized to account for neurons that admit persistent mixed selectivity. This shows that, besides modular organization, the model can also incorporate centralized hierarchical organization which is a necessary ingredient for integration~\cite{shanahan2010broadcast,zamora2010cortical,zamora2011exploring,sporns2013network}. In a previous study, based on phase oscillator models coupled via gap junctions and in presence of phase-difference-dependent plasticity (PDDP), the mixed selectivity was only a transient phenomenon. Indeed in Ref.~\cite{bergoin2023inhibitory}, the PDDP tends to merge the hubs to one of the shared memories in the long-term, while in the spiking network with STDP here considered, the hubs are stable features of the connectivity structure.

\section*{Summary and Discussion}

The architecture of brain's connectivity, the dynamics of neural populations and the learning capacities of neural networks are three fundamental topics of computational neuroscience. However, these topics are too often investigated separately one from another. In line with the research presented in~\cite{litwin2014,zenke2015,ocker2019,yang2024}, our aim here was to develop a learning model that exhibits the emergence of relevant architectural features but which remains \emph{alive} in all phases. Meaning that neither the dynamical activity nor the plasticity evolution were frozen once the training is finalized. This is also referred as continuous learning. The model displays realistic firing patterns after the learning phase while the synaptic plasticity remains active---as it is the case in the brain \emph{in-vivo}.

To achieve these objectives, we have introduced a network of excitatory and inhibitory QIF neurons with plastic synapses following simple STDP rules based only on local information, i.e. the pre- and post-synaptic spike times~\cite{bi1998synaptic}. By targeting stimuli to distinct sub-populations---mimicking the segregated projections of different features into early sensory layers---the network developed a stable modular connectivity. Moreover, by allowing the stimuli to target also neurons in overlapping sub-populations, we observed the emergence of hub-like neurons displaying mixed selectivity. Furthermore, the learned connectivity patterns were robust against catastrophic forgetting over time despite plasticity remained active after the learning phase, and in the absence of any homeostatic or control mechanisms, in contrast to previous analyses~\cite{litwin2014,zenke2015,ocker2019,yang2024}. As a matter of fact, we found that the key factor to guarantee the reinforcement and the long-term maintenance of the memories was the spontaneous occurrence of transient memory recalls in the post-learning neural dynamics. These events occurred at random times on top of an otherwise asynchronous and irregular background neuronal activity. In particular, the recalls acted as short, punctuated boosts to the synaptic adaptation which allowed the memory patterns to persist against natural forgetting.

\subsection*{The role of feedback and feedforward inhibition}

Inhibition has often been considered to play a mere control function by avoiding pathological situations characterized by extremely high firing activity~\cite{mi2017synaptic,taher2020exact}. In our model, instead, inhibition also plays an active role in the formation and stabilization of the memories. Consequently, the memory capacity of the model not only depends on the number of excitatory neurons, but also on the number of inhibitory ones.
Although inhibitory neurons account for only around 20\% of neural cells in the human brain, GABAergic interneurons represent the majority of sub-classes present in the brain~\cite{petilla2008petilla}. There is substantial evidence that inhibitory cells are regulated by different types of synaptic plasticity~\cite{kullmann2012plasticity,koch2013hebbian} promoting, for example, the creation of feed-forward~\cite{lamsa2005hebbian,wu2022regulation} and feedback inhibitory circuits~\cite{lamsa2007anti,wu2022regulation}. 
However, whether a direct relation exists between the type of plasticity and the various functionalities of inhibitory neurons is yet largely to be clarified~\cite{guy2023direction}. Accordingly, we opted for exploring the behaviour of the model under Hebbian and anti-Hebbian inhibition. 

During the training, the Hebbian inhibitory neurons formed internal feedback loops (with the excitatory neurons targetted by the same stimulus, Figs.~\ref{fig:Figure1}C,D) while the anti-Hebbian neurons formed lateral feed-forward connections (across the populations that were targetted by different stimuli, Figs.~\ref{fig:Figure1}B,D). As a consequence, the Hebbian feedback inhibition turned responsible for controlling the firing rate of the populations, which is crucial for the stabilization of the network activity into a realistic asynchronous irregular dynamics~\cite{vogels2011inhibitory,politi2024} and to prevent abnormal behaviour, e.g., the pathological high frequency firing observed in Fig.~\ref{fig:Figure1}B. Meanwhile, the anti-Hebbian feed-forward inhibition mediates the competition and selective activation across populations by, e.g., allowing the stimulated population (the stored memory item) to silence the activity of the other populations, representing other memory items. 

These results resonate with recent empirical and modeling observations. Lagzi, et al.~\cite{lagzi2021assembly} showed via \emph{in-vitro} experiments that parvalbumin-expressing (PV) and somatostatin-expressing (SOM) interneurons of the mouse frontal cortex follow symmetric and asymmetric Hebbian STDP respectively. Subsequent modeling suggested PV neurons to mediate homeostasis in excitatory activity, while SOM neurons build lateral inhibitory connections providing competition between excitatory assemblies. Guy et al.~\cite{guy2023direction} performed \emph{in-vivo} measurements of the responses of PV, SOM and vasoactive intestinal peptide-expressing (VIP) interneurons at the mouse barrel cortex and found differentiated angular selectivity functionality for each neuron type. In particular, whisker stimulation evoked direction-selective inhibition in a majority of SOM interneurons.

Previous models have compared the function of Hebbian and anti-Hebbian plasticities but these were investigated either separately~\cite{foldiak1990forming,vogels2011inhibitory,luz2012balancing,kleberg2014excitatory} or assuming that one inhibitory neuron can be susceptible for both plasticity types~\cite{zenke2015}. Here, we considered two populations of inhibitory neurons---one population subject to Hebbian plasticity and the other to anti-Hebbian---and thus, we studied their joint impact on the adaptation and the dynamical behaviour of a neuronal network, allowing us to naturally associate both types of plasticity with given functions. In our model the feedback and feed-forward inhibitory circuits emerged spontaneously---through adaptation---and were not imposed a priori. In the light of our results, we conclude that the coexistence of Hebbian and anti-Hebbian inhibition is decisive for the formation of stable memory-related structural modules (assemblies) and for the onset of spontaneous memory recalls in the neuronal activity.

\subsection*{The relevance of spontaneous recalls}

The consolidation and long-term preservation of memories involve neurochemical changes that are modulated by the dynamical activity of the neurons and their neighbours~\cite{bliss1993synaptic,mcgaugh2000memory,malenka2004ltp}. Cortical and hippocampal activity during REM and non-REM sleep is characterized by low and high frequency firing events~\cite{torao2021up,wang2022no} that are relevant for the consolidation of daytime experiences~\cite{squire1995retrograde,steriade2001impact,stickgold2005sleep,marzano2011recalling,dudai2012restless,theodoni2018theta,eichenlaub2014resting}. The cortical activity during resting awake is typically characterised by asynchronous irregular dynamics, facilitating the integration of various sensory inputs and the formation of new associations between different information sources~\cite{buzsaki2004neuronal,fries2005mechanism}. Additionally, short and random events of partial synchronous activation---occurring on top of the irregular firing background---are associated to spontaneous memory recalls (or retrieval)~\cite{gu2016dynamics} and to the consolidation~\cite{vogels2011inhibitory,carrillo2016imprinting,lagzi2021assembly} of learned memories.
 
Once the learning phase has finalized, the dynamical behaviour of our model resembles that of awake resting-state \emph{in-vivo} (Figs.~\ref{fig:Figure2}C-J). It shows both a persistent asynchronous irregular activity with low firing ($\approx 0.0-5$ Hz) and punctuated events of transient synchrony of higher rates ($\approx 8-15$ Hz). The STDP rules are effective models of the relation between the dynamical activity of neurons and the underlying biochemical processes leading to synaptic plasticity~\cite{stdp}. STDP stresses that these biochemical changes---either for depression or potentiation---take effect primarily when neurons fire concurrently within short time windows. In consequence, if the activity of a group of neurons is uncorrelated over time, the memories they form tend to be slowly forgotten. 
In our model, this slow forgetting is compensated by the occurrence of spontaneous (partial or complete) recalls of the previously learned memories, happening at random times. These recalls act as short boosts reinforcing the patterns of synaptic connectivity formed during the learning, and thus constitute an autonomous mechanism for memory consolidation during rest. These results emphasize the importance of considering dynamical models of learning which remain \emph{alive} after training~\cite{litwin2014,zenke2015,ocker2019,yang2024}---instead of freezing their activity and plasticity---in order to understand the processes leading to memory consolidation and long-term maintenance.

It shall also be stressed that the coexistence of asynchronous irregular firing and memory recalls is an emergent dynamical property of our model, instead of being an enforced behaviour by the addition of internal mechanisms for frequency adaptation or by the application of external modulation. The asynchronous irregular activity emerges in our globally coupled model thanks to the three following: presence of non-identical neurons mimicking different neural properties, distributed synaptic weights, and external noise. However, analogous results can be obtained in randomly connected deterministic networks due to internal self-generated fluctuations as shown in Supplementary Sec.~\ref{sec:supp6}, as well as in previous analyses~\cite{litwin2014,zenke2015,yang2024}. In a previous study with phase oscillators~\cite{bergoin2023inhibitory}, the post-learning activity at rest showed a partially synchronous state similar to a limit cycle. Here the network remains stable in its asynchronous regime while the recalls spontaneously \emph{pop out} at random times and order, allowing more numerous and complex information to be encoded and maintained. This behaviour naturally results from the interplay between the underlying structural organization---shaped by the learning process---and the background activity eliciting the transitions, somehow similar to the noise-driven state switching found in neural activity~\cite{rolls2010noisy}. These results evidence the benefit of studying connectivity, dynamics and learning simultaneously, as they represent different faces of interdependent phenomena.

\subsection*{The memory capacity of the model}

Empirical and theoretical studies have associated inhibitory plasticity with the long-term storage of memories~\cite{karunakaran2016PVplasticity,barron2017inhibitory,mongillo2018inhibitory,giorgi2021roles}. From these indications one could deduce the idea that the (long-term) memory capacity of the brain might be related to the quantity of inhibitory neurons. We have shown that the memory capacity of our model depends on the number of both excitatory and inhibitory neurons.
The maximal storage capacity of our model is $M^\star = 0.333 N$ memory items (Fig.~\ref{fig:Figure4}B), where $N$ is the total number of neurons. This represents more than twice the memory capacity of the classical Hopfield model ($\simeq0.14N$~\cite{hopfield1982neural,amit1985}) which does not respect the Dale’s principle---it allows the connections to take either positive or negative values, thus ignoring the separation of neurons into excitatory and inhibitory classes. It should be stressed that the value $M^\star$ should be considered as a theoretical limit. It arises from the fact that, in our model, each memory needs to be associated to at least three neurons: a triplet formed by one excitatory, one Hebbian inhibitory and one anti-Hebbian inhibitory neuron. While in correspondence of this limit values the stability of all the memories is guaranteed, achieving this capacity might be rather implausible for biological neural networks. For example, a real brain operating at this limit would be very fragile since the lesion of a single neuron would be sufficient to erase a memory. Also, this limit requires the $E/I$ ratio to be $33/66$, while in the human cortex this is known to be $80/20$. For this ratio, the maximal capacity of our model is indeed $0.1 N$.

An overview of the known variability of $E/I$ ratios across different parts of the human brain (Table in Fig.~\ref{fig:Figure4}C) reveals that smaller regions such as the amygdala, thalamus, striatum and globus pallidus tend to have a higher proportion of inhibitory neurons than the cortex or the cerebellum. Extrapolating the results for the memory capacity from our model, it allows us to speculate that regions with fewer resources for memory storage (in terms of total number of neurons), may compensate by allowing for a more favourable $E/I$ ratio. Similarly, the mouse admits a more favourable ratio than humans in some areas such as the basal ganglia, striatum, amygdala or the hippocampus. While this interpretation is at the moment largely hypothetical, it may reflect a biological strategy for resource optimisation. 
Furthermore, it shall be noted that memory capacity is further affected by the complexity of the memories; with simple ones requiring less neurons to be stored and complex memories each requiring more neurons. Hence, the empirically observed $E/I$ ratios may also reflect a compromise between the number of memories and the complexity of the information that needs to be stored.

Real neurons generally respond to---and encode for---multiple inputs and memories~\cite{quiroga2005}. This mixed selectivity plays a crucial role in complex cognitive tasks allowing the brain to simultaneously represent and integrate multiple sources of information~\cite{rigotti2013importance,johnston2020nonlinear}. In addition to retaining segregated memories, our model also exhibits the possibility of learning and recalling complex memories admitting mixed selectivity, see Fig.~\ref{fig:Figure6}. The neurons supporting mixed selectivity can be considered as hub neurons~\cite{bonifazi2009} connecting between the stored clusters and facilitating the ability to transmit and integrate information~\cite{zamora2010cortical,van2011rich}. The presence of hub neurons does not increase the memory capacity of a network but instead, they allow for more complex information to be encoded by establishing associations between memories.

\subsection*{Limitations and outlook}

Despite the model was developed to respect several biological constrains and to satisfy some realistic behavioural aspects, other choices and limitations could be implemented in future refinements. We omitted possible evolutionary aspects and assumed that only synaptic plasticity to be responsible for shaping the structure of the connectivity. Sensory systems organize information spatially with neighbouring neurons representing similar response properties and forming, e.g., retinotopic, tonotopic or somatotopic maps~\cite{mountcastle1957modality,kaas1997topographic,kaas2001organization}. However, in the present model the neurons form a graph with no spatial embedding. It would be thus a natural step to extend the model in the future with spatially distributed neurons, following biologically representative spatial constraints.

The long-range white matter connections between distant cortical regions are mainly formed by axonal bundles of excitatory pyramidal neurons~\cite{petreanu2009subcellular,zhang2014long}, although some GABAergic cells have also been found to project across different brain regions~\cite{bonifazi2009,caputi2013long}. 
When our model was trained with stimuli targetting separate neuronal populations, excitatory neurons formed local connections and only the anti-Hebbian inhibitory neurons developed ``long-range'' projections to other clusters. While this architecture is consistent with the lateral inhibition found in circuits for selectivity and decision-making~\cite{gerstner2014,maass2000computational,guy2023direction}, it is not representative of the excitatory long-range white matter fibers in the cortex. However, in the model, cross-modular excitatory connections appeared when the stimuli acted on overlapping populations, leading to the formation of excitatory hubs. While the reasons for the organization of excitatory and inhibitory cross-modular connectivities requires further investigation, extensions of the model here presented might be relevant to explore those mechanisms. In fact, both the organization of micro-/macroscopic connectivities and the development of short-/long-term memories, might be governed by distinct rules which could be separately implemented in the model.

\section*{Methods}
\label{sec:methods}

This section describes the spiking neuronal network model governing the dynamics of the neurons and the learning rules used for the adaptation of synaptic weights, as well as the microscopic and macroscopic indicators employed to characterize the network states and dynamics in the paper.

\subsection*{Spiking neuronal network model}

We consider a network of QIF neurons, pulse coupled via exponentially decaying post-synaptic potentials and in the presence of STDP. Unless otherwise specified, the network will be composed of 80\% (20\%) excitatory (inhibitory) neurons as usually observed in the human cortex~\cite{abeles1991corticonics}. Depending on the plasticity rules controlling the synaptic strengths of the connections, three neural sub-populations can be identified depending on the nature of the pre-synaptic neurons:
\begin{itemize}
	\item excitatory neurons subject to asymmetric Hebbian STDP;
	\item inhibitory neurons subject to symmetric Hebbian STDP;
	\item inhibitory neurons subject to symmetric anti-Hebbian STDP.
\end{itemize}
The evolution of the membrane potential $V_{i}$ of each neuron ($i = 1,...,N$) is described by the following ordinary differential equation:
\begin{equation}
	\label{eq:1}
	\tau_{m} \dot{V_{i}} = V_{i}^{2}(t) + \eta_{i} + g_{e} S_{i}^{e}(t) + g_{hi} S_{i}^{hi}(t) + g_{ai} S_{i}^{ai}(t) + I_{i}(t) + \xi_{i}(t) ,
\end{equation}
where $\tau_{m} = \tau_0$ ms (with $\tau_0=20$) is the membrane time constant; $\eta_{i} \sim \mathcal{N}(0.0,\, (\pi \tau_{0})^{2})$ are the neuronal excitabilities, chosen to have an average neuronal firing rate at rest of around $1$ Hz, $I_{i}(t) = \{0, (50 \pi \tau_{0})^{2}\}$ are the external DC currents, leading the neurons to fire around $50$ Hz whenever stimulated and $\xi_{i}(t) \sim \mathcal{N}(0.0,\, (4 \pi \tau_{0})^{2})$ is a Gaussian white noise term tuned to induce a firing variability of $\simeq 4$ Hz. 
Whenever the membrane potential $V_i$ reaches infinity, a spike is emitted and $V_i$ is reset to $-\infty$. In the absence of synaptic coupling, external DC current and noise, the QIF model displays excitable dynamics for $\eta_i < 0$, while for positive $\eta_i$, it behaves as an oscillator with period $T_i^{(0)}/\tau_m= \pi/\sqrt{\eta_i}$~\cite{ermentrout1986}. 

The synaptic dynamics is mediated by the global synaptic strengths : $g_{e}=100$, $g_{hi}=400$ and $g_{ai}=200$ for the excitatory, the Hebbian and anti-Hebbian inhibitory neurons, respectively. Finally, the evolution of the excitatory, Hebbian inhibitory and anti-Hebbian inhibitory synaptic currents $S_{i}^{e}(t)$, $S_{i}^{hi}(t)$ and $S_{i}^{ai}(t)$ is given by 
\begin{equation}
	\label{eq:2}
	\tau_{d}^{e} \dot{S_{i}^{e}} = - S_{i}^{e} + \dfrac{\tau_{d}^{e}}{N_{e}} \sum_{j=1}^{N_{e}}
	\sum_n w_{ij}(t) \delta(t-t_{j}^{(n)}) ,
\end{equation}
\begin{equation}
	\label{eq:3}
	\tau_{d}^{i} \dot{S_{i}^{hi}} = - S_{i}^{hi} + \dfrac{\tau_{d}^{i}}{N_{hi}} \sum_{j=1}^{N_{hi}} 
	\sum_n w_{ij}(t) \delta(t-t_{j}^{(n)}) ,
\end{equation}
\begin{equation}
	\label{eq:4}
	\tau_{d}^{i} \dot{S_{i}^{ai}} = - S_{i}^{ai} + \dfrac{\tau_{d}^{i}}{N_{ai}} \sum_{j=1}^{N_{ai}} \sum_n 
	w_{ij}(t) \delta(t-t_{j}^{(n)}) ,
\end{equation}
where $\tau_{d}^{e}=2$ ms ($\tau_{d}^{i}=5$ ms) are the exponential time decay for the excitatory (inhibitory) post-synaptic potentials; $w_{ij}(t)$ are the plastic coupling weights from neuron $j$ towards neuron $i$, whose dynamics is described in the following and $t_{j}^{(n)}$ is the $n$-th spike time of the $j$-th pre-synaptic neuron, and $\delta(t)$ is the Dirac delta function.

We consider a network of $N$ all-to-all connected neurons without self-connections and $N = N_{e} + N_{hi} + N_{ai} = 100$, where $N_{e}=80$, $N_{hi}=10$ and $N_{ai}=10$ are the number of excitatory, Hebbian inhibitory and anti-Hebbian inhibitory neurons, respectively.

We integrate Eqs.~\eqref{eq:1}, \eqref{eq:2}, \eqref{eq:3} and \eqref{eq:4} via a stochastic Euler scheme~\cite{toral2014} with time step $dt$ (see Table~\ref{table1} for its value). Whenever $V_{i}(t)$ overcomes a threshold value $V_{p}=10$, we approximate the $n$-th firing time of neuron $i$ as $t^{(n)}_i = t + \frac{1}{V_{i}}\tau_{m}$, where $\frac{1}{V_{i}}\tau_{m}$ corresponds to the time needed to reach $+\infty$ from the value $V_p$. Furthermore, the neuron is reset to $V_{r}=-10$ and held to such value for a time interval $\frac{2}{V_{i}}\tau_{m}$ seconds corresponding to the time needed for the neuron to reach $V_{i}(t)=\infty$ from $V_p$ and to return to $V_r$ after being reset to $-\infty$~\cite{taher2020exact}.

The choice of parameters sets the neurons into a low firing state with a relatively large variability ranging from $0.0$ to $8$ Hz, analogous to those observed in the cortex at rest. 
The external drives $\{I_i(t)\}$ turn the neuronal activity into high frequency oscillations in the $\gamma$ range of approximately $50$ Hz, typical of high-order perceptual activity~\cite{tallon1996stimulus,herrmann2001human}.

\subsection*{STDP rules}

The time evolution of the synaptic weights $w_{ij}(t)$ is controlled by STDP rules which depend on the time difference $\Delta t = t_{i}-t_{j}$ between the last spikes of the post-synaptic neuron $i$ and the pre-synaptic neuron $j$~\cite{bi1998synaptic,stdp,caporale2008,mikkelsen2013}.   

The potentiation of the synapses is controlled by the plasticity function  
\begin{equation}
	\label{eq:8}
	{\Lambda^{+}(\Delta t)} =
	\begin{cases}
		\Lambda(\Delta t) ,&{\text{if}\ \Lambda(\Delta t) \ge 0,} \\
		{0,}&{\text{if}\ \Lambda(\Delta t) < 0,}
	\end{cases}
\end{equation}
while their depression is controlled by
\begin{equation}
	\label{eq:9}
	{\Lambda^{-}(\Delta t)} =
	\begin{cases}
		0,&{\text{if}\ \Lambda(\Delta t) \ge 0,} \\
		{\Lambda(\Delta t) ,}&{\text{if}\ \Lambda(\Delta t) < 0,}
	\end{cases}
\end{equation}
where $\Lambda(\Delta t)$ entering in Eqs.~\eqref{eq:8} and~\eqref{eq:9} depends on the nature of the pre-synaptic neuron.

\begin{figure*}[ht!]
	\centering
	\includegraphics[width=0.85\linewidth]{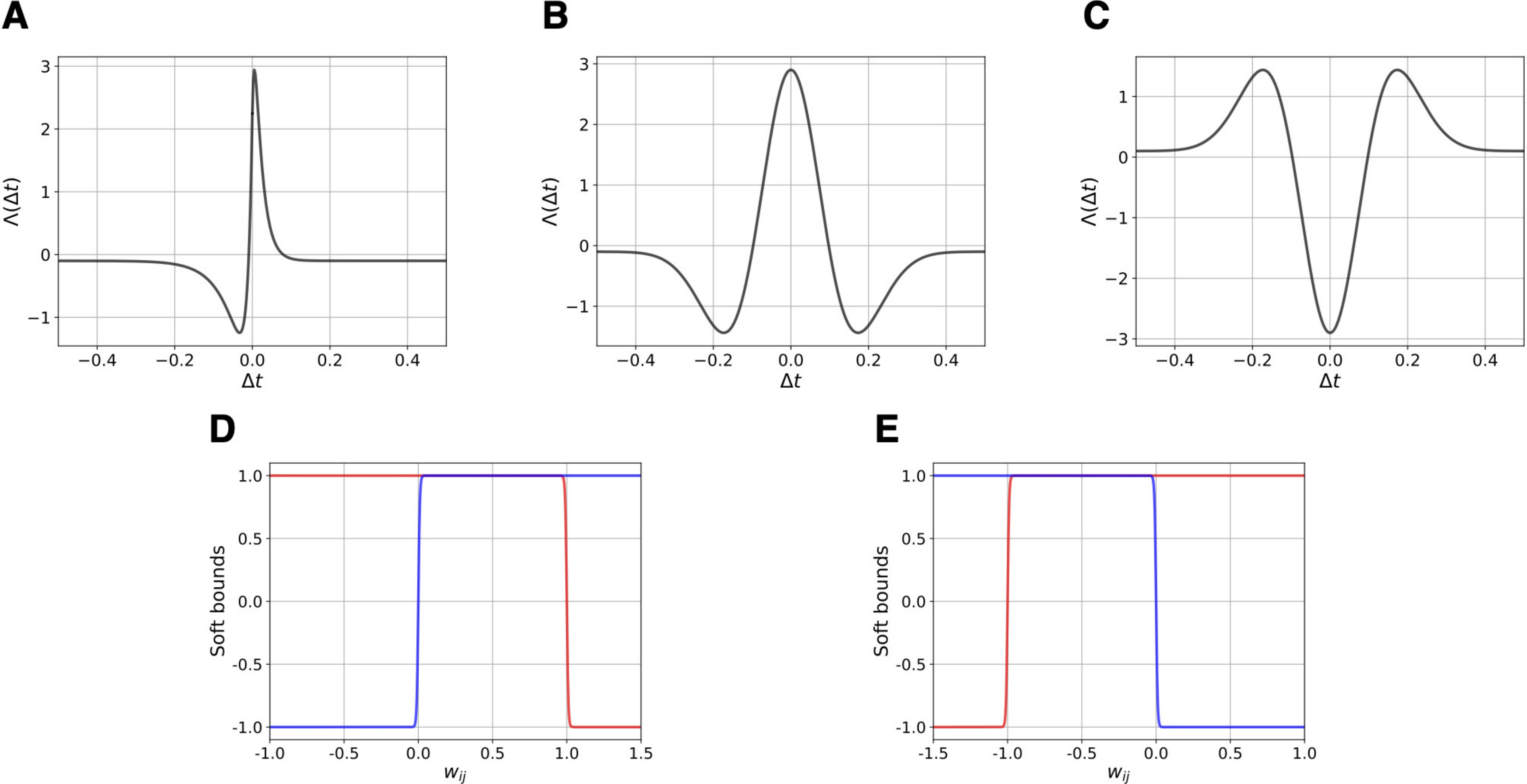}
	\caption{{\bf Plasticity and soft bound functions.}
	Plasticity functions:  {\bf (A)} Hebbian asymmetric STDP $\Lambda_e (\Delta t)$ (Eq.~\eqref{eq:10}); 
	{\bf (B)} Hebbian symmetric STDP $\Lambda_{hi} (\Delta t)$ (Eq.~\eqref{eq:11}); 
	{\bf (C)} anti-Hebbian symmetric STDP $\Lambda_{ai} (\Delta t)$ (Eq.~\eqref{eq:12}). 
	Soft bound functions for excitatory {\bf (D)} and inhibitory {\bf (E)} neurons. The bounds for potentiation are shown in red and for depression in blue. 
	The displayed functions refer to $M=2$ encoded stimuli, i.e. to $f=0.1$.
	} 
	\label{fig:Figure7}
\end{figure*}

For the pre-synaptic excitatory neurons we use an asymmetric Hebbian STDP function $\Lambda_e (\Delta t)$~\cite{bi1998synaptic}. This function affects the evolution of the weights in a causal way: when the pre-synaptic neuron $j$ emits a spike before (after) the post-synaptic neuron $i$, the synaptic weight $w_{ij}$ increases (decreases)~\cite{carlson2013biologically}. The function $\Lambda_e (\Delta t)$ is depicted in Fig.~\ref{fig:Figure7}A, and its explicit expression reads as
\begin{equation}
	\label{eq:10}
	{\Lambda_e (\Delta t)} =
	\begin{cases}
		A_{+} e^{-\frac{\Delta t}{\tau_{+}}} - A_{-} e^{-\frac{4 \Delta t}{\tau_{+}}} - f ,&{\text{for}\ \Delta t \ge 0,} \\
		{A_{+} e^{\frac{4\Delta t}{\tau_{-}}} -A_{-} e^{\frac{\Delta t}{\tau_{-}}} - f,}&{\text{for}\ \Delta t < 0,}
	\end{cases}
\end{equation}
\noindent with the time constants $\tau_{+} = 0.02$ sec and $\tau_{-} = 0.05$ sec, the amplitudes $A_{+} = 5.296$ and $A_{-} = 2.949$. 

The term $f$ models a natural slow forgetting of the memories~\cite{wixted2004psychology,hardt2013decay} by causing a small but constant depression of the weights. We assume that $f$ is inversely proportional to the number $M$ of encoded items: namely, $f = f_0/M$, with $f_0=0.2$. This in order to allow all the stored memory items to be randomly recalled during the post-learning resting phase (for more details, see Supplementary Sec.~\ref{sec:supp9}). Apart from the term $f$, all the other parameters are not modified in the simulations.

For pre-synaptic Hebbian inhibitory neurons, we use a symmetric Hebbian STDP function $\Lambda_{hi} (\Delta t)$~\cite{perez2001hebbian,lamsa2005hebbian,vogels2011inhibitory,luz2012balancing,kleberg2014excitatory,wu2022regulation}, which takes the form of a Ricker wavelet function (or Mexican hat), potentiating (depressing) weights of neurons spiking in a correlated (uncorrelated) way. $\Lambda_{hi} (\Delta t)$ is shown in Fig.~\ref{fig:Figure7}B and it reads as
\begin{equation}
	\label{eq:11}
	\Lambda_{hi}(\Delta t) =
	A \left[1 - \left(\frac{\Delta t}{\tau}\right)^{2}\right] e^{\frac{-\Delta t^{2}}{2\tau^{2}}} - f ,
\end{equation}
\noindent with the time constant $\tau = 0.1$ sec, the amplitude $A = 3$ and the forgetting term $f = 0.1$.

For pre-synaptic anti-Hebbian inhibitory neurons we use a symmetric anti-Hebbian STDP function $\Lambda_{ai} (\Delta t)$ ~\cite{foldiak1990forming,plumbley1993efficient,lamsa2007anti,koch2013hebbian,kleberg2014excitatory,wu2022regulation} which corresponds to a reverse Ricker wavelet function (or reverse Mexican hat), potentiating (depressing) weights of neurons spiking in a uncorrelated (correlated) way. The function $\Lambda_{ai} (\Delta t)$ is shown in Fig.~\ref{fig:Figure7}C and it takes the following expression:  
\begin{equation}
	\label{eq:12}
	\Lambda_{ai} (\Delta t) =
	-A \left[1 - \left(\frac{\Delta t}{\tau}\right)^{2}\right] e^{\frac{-\Delta t^{2}}{2\tau^{2}}} + f ,
\end{equation}
\noindent with the time constant $\tau = 0.1$ sec and the amplitude $A = 3$. In this case, the forgetting term $f = 0.1$ allows to have a constant small potentiation (the rule being anti-Hebbian) of the weights whatever the spike timing difference.

\subsection*{Adaptation of the synaptic weights}

In this sub-section we explain in details the evolution of the synaptic weights. In particular, when the pre-synaptic neuron $j$ or the post-synaptic neuron $i$ spikes at time t, the weight $w_{ij}$ is updated according to the following equations:
\begin{equation}
	\label{eq:5}
	w_{ij}(t^{+}) = w_{ij}(t^{-}) + \gamma_l \Delta w_{ij} ,
\end{equation}
with
\begin{equation}
\label{eq:6}
\Delta w_{ij} =(-1)^{a_q} [\tanh(\lambda(w_q^l - w_{ij})) * \Lambda_{q}^{x(q)}(\Delta t) 
+ \tanh(\lambda (w_{ij}+w_q^u)) * \Lambda_{q}^{y(q)}(\Delta t)] ,
\end{equation}

\noindent where $q$ denotes if the pre-synaptic neuron is excitatory $q=e$ or Hebbian (anti-Hebbian) inhibitory $q=hi$ ($q=ai$). For excitatory (inhibitory) neurons we set $w_q^l=1$ ($w_q^l=0$) and  $w_q^u=0$ ($w_q^u=1$), thus ensuring that the excitatory (inhibitory) couplings are defined within the interval $w_{ij} \in [0:1]$ ($w_{ij} \in [-1:0]$). Moreover, $a_e=2$ and $a_{hi}=a_{ai}=1$ while $x(e)=y(hi)=y(ai) = +$  and $y(e)=x(hi)=x(ai) = -$, thus, for inhibitory synapses, the plasticity functions $\Lambda^{+}(\Delta t)$ and $\Lambda^{-}(\Delta t)$ are exchanged and multiplied by $-1$ since potentiation (depression) of inhibitory weights makes them converge towards $-1$ ($0$). Furthermore, $\gamma_{l} = 0.005$ is the dimensionless learning rate for the adaptation, while the parameter $\lambda=100$ controls the slope of the soft bound function $\tanh(\lambda x)$. 

These non-linear functions, depicted in Figs.~\ref{fig:Figure7}D-E, model the saturation of the synaptic weights $|w_{ij}|$ between $0$ and $1$~\cite{jin2016ap}. The saturation is obtained by allowing for synaptic depression (potentiation) dominating the potentiation (depression) term for large (small) values of the weights~\cite{stdp,van2012soft}. 
The steep slope of these functions (controlled by the parameter $\lambda$) guarantees a dynamical evolution of the synaptic weights even for large values of $w_{ij}$~\cite{gilson2011stability}. In this sense, the functions here used are at the limit between a soft and a hard bound.

We note that the firing activity of the neurons directly impacts the adaption rate and that the synaptic weights are always subject to adaptation, unlike in conventional artificial neural networks. This differs from the previous study with phase oscillators and PDDP~\cite{bergoin2023inhibitory}, where two different learning timescales were needed to allow for the storing and maintenance of the memory items.

All network parameters employed throughout this study are summarized in Table~\ref{table1}.

\begin{table}[!ht]
	\centering
	\caption{\bf Parameters for the network of QIF neurons.}
	\begin{tabular}{| c | c |}
		\hline
		\bf Parameters & \bf Values \\ 
		\hline
		$N$ & $100$ \\
		\hline
		$N_{e}$ & $80$ \\
		\hline
		$N_{hi}$ & $10$ \\
		\hline
		$N_{ai}$ & $10$ \\
		\hline
		$g_{e}$ & $100$ \\
		\hline
		$g_{hi}$ & $200$ \\
		\hline
		$g_{ai}$ & $400$ \\
		\hline
		$\eta$ & $\mathcal{N}(0.0,\, (\pi \tau_{0})^{2})$ \\
		\hline
		$I(t)$ & $ \{0, (50 \pi \tau_{0})^{2}\}$ \\
		\hline
		$\xi(t)$ & $ \mathcal{N}(0.0,\, (4 \pi \tau_{0})^{2})$ \\
		\hline
		$V_{p}$ & $10$ \\
		\hline
		$V_{r}$ & $-10$ \\
		\hline
		$\tau_{0}$ & $0.02$  \\
		\hline
		$\tau_{m}$ & $0.02$ sec \\
		\hline
		$\tau_{d}^{e}$ & $0.002$ sec \\
		\hline
		$\tau_{d}^{i}$ & $0.005$ sec \\
		\hline
		$\tau_{+}$ & $0.02$ sec \\
		\hline
		$\tau_{-}$ & $0.05$ sec \\
		\hline
		$\tau$ & $0.1$ sec \\
		\hline
		$dt$ & $0.001$ sec \\
		\hline
		$\gamma_l$ & $0.005$ \\
		\hline
		$A_{+}$ & $5.296$ \\
		\hline
		$A_{-}$ & $2.949$ \\
		\hline
		$A$ & $3$ \\
		\hline
		$f$ & $0.1$ \\
		\hline
		$\lambda$ & $100$ \\
		\hline
	\end{tabular}
	\label{table1}
\end{table}

\subsection*{Microscopic and macroscopic indicators}

In this sub-section, we define the indicators employed to characterize the network dynamics at a microscopic and macroscopic level. \\

\paragraph*{Microscopic Indicators}
The behaviour of single neurons was quantified by their instantaneous firing rates $\nu_j(t)$ defined as
\begin{equation}
	\label{eq:15}
	\nu_{j}(t) = \frac{n_{j}^{sp}(t)}{T} \; .
\end{equation}
Here, $n_{j}^{sp}(t)$ is the number of spikes emitted by neuron $j$ (i.e. the spike count) in the time interval  $[t:t+T]$, with $T=0.05$ sec. 

For a population of $N_{p}$ neurons, their instantaneous activity can be characterized as follows
\begin{equation}
	\label{eq:16}
	\nu_p(t) = \frac{1}{N_{p}} \sum_{j=1}^{N_{p}} \nu_{j} (t) \; .
\end{equation}
 
The firing rate variability has been measured in terms of the coefficient of variation $CV_j = \frac{\sigma_j}{\mu_{j}}$ where $\mu_j$ is the mean interspike interval (ISIs) of the neuron $j$ and $\sigma_j$ its standard deviation. For a perfectly periodic firing $CV_j=0$, while for a Poissonian process $CV_j=1$.  \\

\paragraph*{Macroscopic Indicators}
The degree of synchronization in the network was quantified by the complex Kuramoto order parameter~\cite{kuramoto2003chemical} 
\begin{equation}
\label{eq:13}
Z(t) = R(t) {\rm e}^{i \Phi(t)} = \frac{1}{N} \sum_{j=1}^{N} {\rm e}^{i \theta_{j}(t)} \, ,
\end{equation}
where $R(t)$ ($\Phi(t)$) represents the modulus (phase) of the macroscopic indicator.
The modulus $R$ is employed to characterize the level of phase synchronization in the network: $R > 0$ ($R=1$) for a partially (fully) synchronized network, while $R \simeq {\cal O}(1/\sqrt{N})$ for an asynchronous dynamics due to finite size effects.

To associate a continuous phase $\theta_{j}(t) \in [0:2 \pi]$ to the spiking activity of neuron $j$, we proceed in the following way:
\begin{equation}
	\label{eq:14}
	\theta_{j}(t) = 2 \pi \frac{ (t-t_{j}^{(n)})}{(t_{j}^{(n+1)}-t_{j}^{(n)})} \qquad t_j^{(n)} \leq t \leq t_j^{(n+1)},
\end{equation}
with $t_{j}^{(n)}$ the $n$-th firing time of neuron $j$.

The mean rate of variation of the synaptic weights was estimated as
\begin{equation}
	\label{eq:17}
	K(t) = \frac{1}{N*(N-1)} \sum_{i=1}^{N} \sum_{j \ne i}^{N} \frac{[ w_{ij}(t+\Delta t)-w_{ij}(t)]}{\Delta t} \; ,
\end{equation} 
where $w_{ij}(t)$ and $w_{ij}(t+\Delta t)$ are the synaptic coupling weights from neuron $j$ to $i$ at times $t$ and $t+\Delta t$ respectively, $\Delta t = 0.1$ sec.
The normalization term is $N*(N-1)$ since we consider an all-to-all connected network without autapses. The parameter $K(t)$ takes positive (negative) values for an overall increase (decrease) in the weight connectivity.

\section*{Data Availability}

All code written in support of this publication is publicly available at \url{https://github.com/rbergoin/QIF-neurons-with-3-synaptic-plasticities/}.

\acknowledgements{
This work was supported (R.B., G.D. and G.Z.L.) by the European Union's Horizon 2020 Framework Programme for Research and Innovation under the Specific [Grant Agreement No. 945539 (Human Brain Project SGA3)] and by an EUTOPIA funding [EUTOPIA-PhD-2020-0000000066 - NEUROAI]. A.T. received financial support by the Labex MME-DII [Grant No. ANR-11-LBX-0023-01] (together with M.Q.), and by the ANR Project ERMUNDY [Grant No. ANR-18-CE37-0014] all part of the French program ``Investissements d'Avenir''. G.D. is supported by the Spanish national research project [ref. PID2019-105772GB-I00/AEI/10.13039/501100011033] funded by the Spanish Ministry of Science, Innovation, and Universities (MCIU). M.Q. is also partially supported by CNRS through the IPAL lab in Singapore. The authors thank Matthieu Gilson for useful discussions.
} 


\clearpage
\newpage
\widetext


\setcounter{equation}{0}
\setcounter{figure}{0}
\setcounter{table}{0}
\setcounter{page}{1}
\setcounter{section}{0}
\makeatletter

\renewcommand{\theequation}{S\arabic{equation}}
\renewcommand{\thefigure}{S\arabic{figure}}
\renewcommand{\thesection}{S\arabic{section}}
\renewcommand{\bibnumfmt}[1]{[S#1]}
\renewcommand{\citenumfont}[1]{S#1}

\begin{flushleft}
	{\Large
		Supplementary Information for: \\
		\vspace{0.5cm}
		\textbf{Emergence and maintenance of modularity in neural networks with Hebbian and anti-Hebbian inhibitory STDP}
	}
	\newline
	\newline
	Rapha\"el Bergoin\textsuperscript{1,2,3,4*},
	Alessandro Torcini\textsuperscript{5},
	Gustavo Deco\textsuperscript{2,3,6},
	Mathias Quoy\textsuperscript{1,7},
	Gorka Zamora-L\'opez\textsuperscript{2,3}
	\\
	\bigskip
	\textbf{1} ETIS, UMR 8051, ENSEA, CY Cergy Paris Universit\'e, CNRS, 6 Av. du Ponceau, 95000 Cergy-Pontoise, France.
	\\
	\textbf{2} Center for Brain and Cognition, Pompeu Fabra University, Ram\'on Trias Fargas 25-27, 08005 Barcelona, Spain.
	\\
	\textbf{3} Department of Information and Communication Technologies, Pompeu Fabra University, Roc Boronat 138, 08018 Barcelona, Spain.
	\\
	\textbf{4} Institute of Neural Information Processing, Center for Molecular Neurobiology (ZMNH), University Medical Center Hamburg-Eppendorf (UKE), 20251 Hamburg, Germany.
	\\
	\textbf{5} Laboratoire de Physique Th\'eorique et Mod\'elisation, UMR 8089, CY Cergy Paris Universit\'e, CNRS, 2 Av. Adolphe Chauvin, 95032 Cergy-Pontoise, France.
	\\
	\textbf{6} Instituci\`{o} Catalana de Recerca i Estudis Avan\c{c}ats (ICREA), Passeig Lluis Companys 23, 08010 Barcelona, Spain.
	\\
	\textbf{7} IPAL, CNRS, 1 Fusionopolis Way \#21-01 Connexis (South Tower), Singapore 138632, Singapore.
	\\
	\bigskip
	* raphael.bergoin@gmail.com
\end{flushleft}

\vspace{2cm}

\clearpage
\section{Untrained group of neurons.} 
\label{sec:supp1}

\begin{figure}[hb!]
	\centering
	\includegraphics[width=0.85\linewidth]{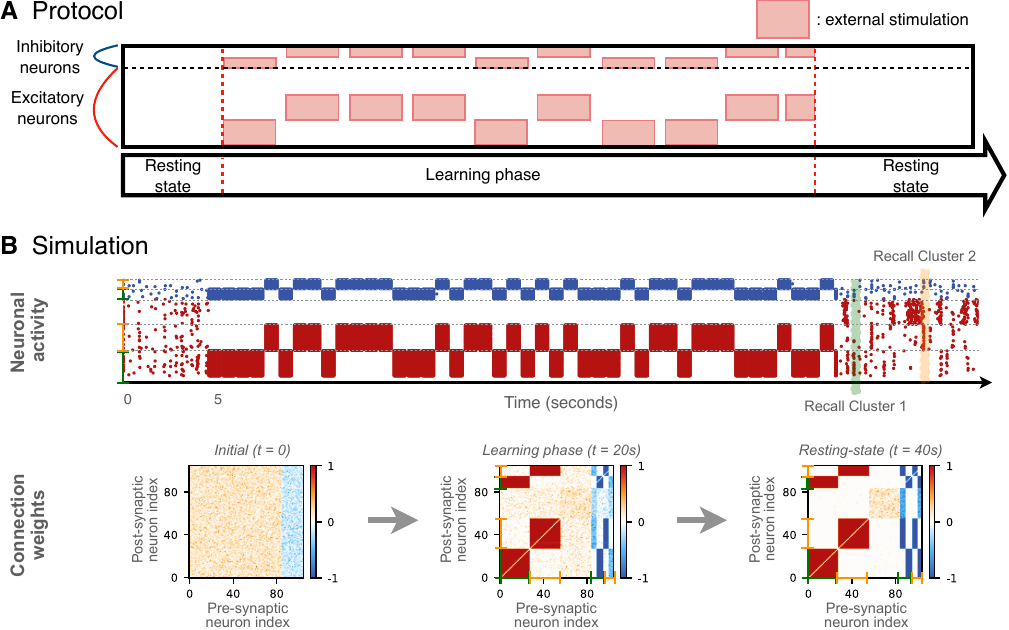}
	\caption{
		{\bf Learning of 2 stimuli with an untrained sub-population.} 
		{\bf (A)} Stimulation protocol for a network of $N=105$ neurons entrained with $M=2$ stimuli with an untrained group of excitatory neurons. 
		{\bf (B)} Simulation and learning results. Connectivity matrices show the evolution of the synaptic weights leading to the emergence of two modules and the decoupling of the unstimulated sub-population. The raster plot shows the simulation for the three stages: initial resting phase, entrainment stage and the post-learning neuronal activity characterized by spontaneous recall events of P$_1$ neurons (green shadow) and P$_2$ neurons (orange shadow).	
	}
	\label{fig:S1Appendix}
\end{figure}

This alternative protocol is analogous to the one of the numerical experiment reported in Fig.~\ref{fig:Figure1}D of the main text. The only difference lies in the fact that a group of excitatory neurons is never stimulated and so it is untrained. The results obtained are described in Fig.~\ref{fig:S1Appendix}. On the one hand, we obtain analogous results with the formation of two modular structures in the weighted connectivity associated to spontaneous recalls of the two different memories during the dynamical evolution as shown in raster plot. On the other hand, neurons of the untrained group are weakly connected among them in accordance with the absence of stimulation and are decoupled from the other clusters while receiving anti-Hebbian inhibition from them. As a result, these neurons spike in a totally asynchronous and irregular way, without impacting the dynamics of the rest of the network.

\vspace{1.5cm}

\clearpage
\section{Randomly stimulated neurons within each population.} 
\label{sec:supp2}

\begin{figure}[hb!]
	\centering
	\includegraphics[width=0.85\linewidth]{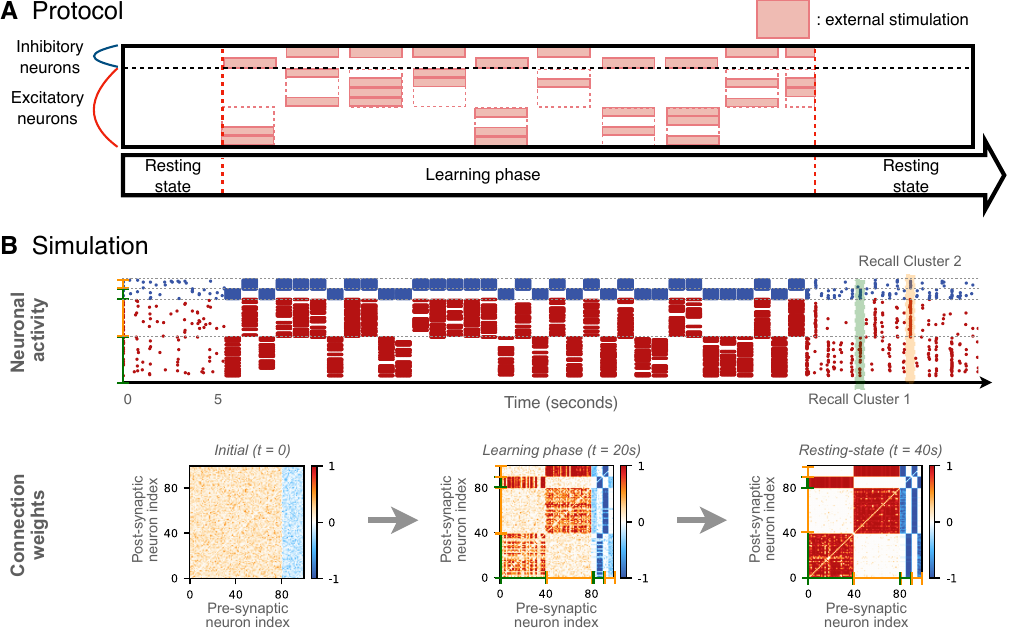}
	\caption{
		{\bf Learning of 2 stimuli with randomly stimulated neurons within each distinct population.} 
		{\bf (A)} Stimulation protocol for a network of $N=100$ neurons entrained with $M=2$ stimuli, neurons within each population being randomly selected. 
		{\bf (B)} Simulation and learning results. Connectivity matrices show the evolution of the synaptic weights leading to the emergence of two modules. The raster plot shows the simulation for the three stages: initial resting phase, entrainment stage and the post-learning neuronal activity characterized by spontaneous recall events of P$_1$ neurons (green shadow) and P$_2$ neurons (orange shadow).	
	}
	\label{fig:S2Appendix}
\end{figure}

This alternative protocol is analogous to that of the numerical experiment shown in Fig.~\ref{fig:Figure4}D of the main text. The only difference lies in the fact that when a population is selected during learning, a random number of neurons in the excitatory population (with a probability of $0.5$) is stimulated. The results obtained are described in Fig.~\ref{fig:S2Appendix}. The direct consequence is that the two modular structures in the weighted connectivity are less well formed compared to the original experiment. Nevertheless, the clusters remain decoupled with the same feedforward and feedback inhibition described in the main text. Only the weights within the clusters appear have more random values. Therefore, although the spontaneous recalls of the two memory items are present in the dynamics as shown in the raster plot, they appears to be somewhat sparser and less synchronized. We assume that this is largely due to the fact that the connections to and from the inhibitory neurons are incomplete. In Fig.~\ref{fig:Figure3}A of the main text, the randomness of the E-E connections does not impact the spontaneous recalls. This highlights the need to reach a convergence of the weights linked to inhibition to correctly memorize the items. Nevertheless, this experiment also shows that even if memory items are partially learned during each period of stimulation, the entire original memory item is somehow retrieved and learned.

\vspace{1.5cm}

\clearpage
\section{Random stimulation values.} 
\label{sec:supp3}

\begin{figure}[hb!]
	\centering
	\includegraphics[width=0.85\linewidth]{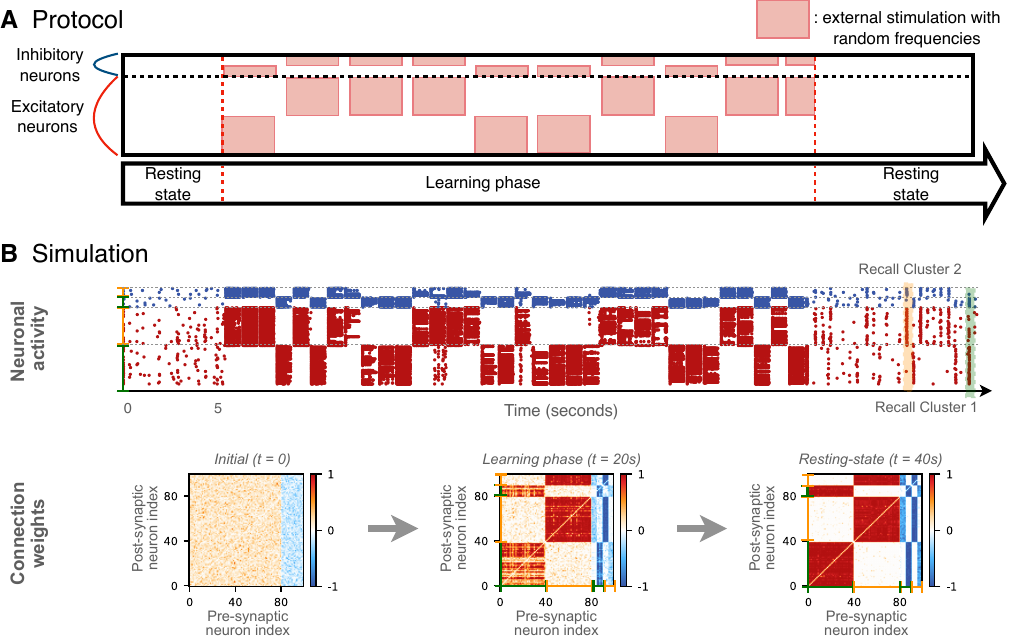}
	\caption{
		{\bf Learning of 2 stimuli with random frequencies.} 
		{\bf (A)} Stimulation protocol for a network of $N=100$ neurons entrained with $M=2$ stimuli of random amplitude.
		{\bf (B)} Simulation and learning results. Connectivity matrices show the evolution of the synaptic weights leading to the emergence of two modules. The raster plot shows the simulation for the three stages: initial resting phase, entrainment stage and the post-learning neuronal activity characterized by spontaneous recall events of P$_1$ neurons (green shadow) and P$_2$ neurons (orange shadow).
	}
	\label{fig:S3Appendix}
\end{figure}

This alternative protocol is again analogous to that of experiment of Fig.~\ref{fig:Figure1}D of the main text. The only difference lies in the fact that when a population is stimulated (excitatory and inhibitory neurons), the neurons within it receive inputs of random amplitude (i.e. inducing firing activity between 50 and 100 Hz). The results obtained are described in Fig.~\ref{fig:S3Appendix}. We observe very similar results to those in Fig.~\ref{fig:S2Appendix}. However in this case, the weight connections seem stronger than in the previous experiment. As a result, the spontaneous recalls appear to be more visible in the dynamics of raster plot. In addition to the conclusions made in the previous study, this experience allows us to conclude that the spatio-temporal correlations of the applied inputs are more impactful on the formation of the modular structures than the intensities of these same inputs. Indeed, the firing frequency induced by the inputs applied to the neurons necessarily has an impact on the encoding of information and the construction of memory as can be seen here. Nevertheless, compared with the previous experiment where the neurons received the same inputs but at times that were not necessarily correlated, the structure is much better learned in the current case. 

\vspace{1.5cm}

\clearpage
\section{Stability of four structural modules.} 
\label{sec:supp4}

\begin{figure}[hb!]
	\centering
	\includegraphics[width=0.85\linewidth]{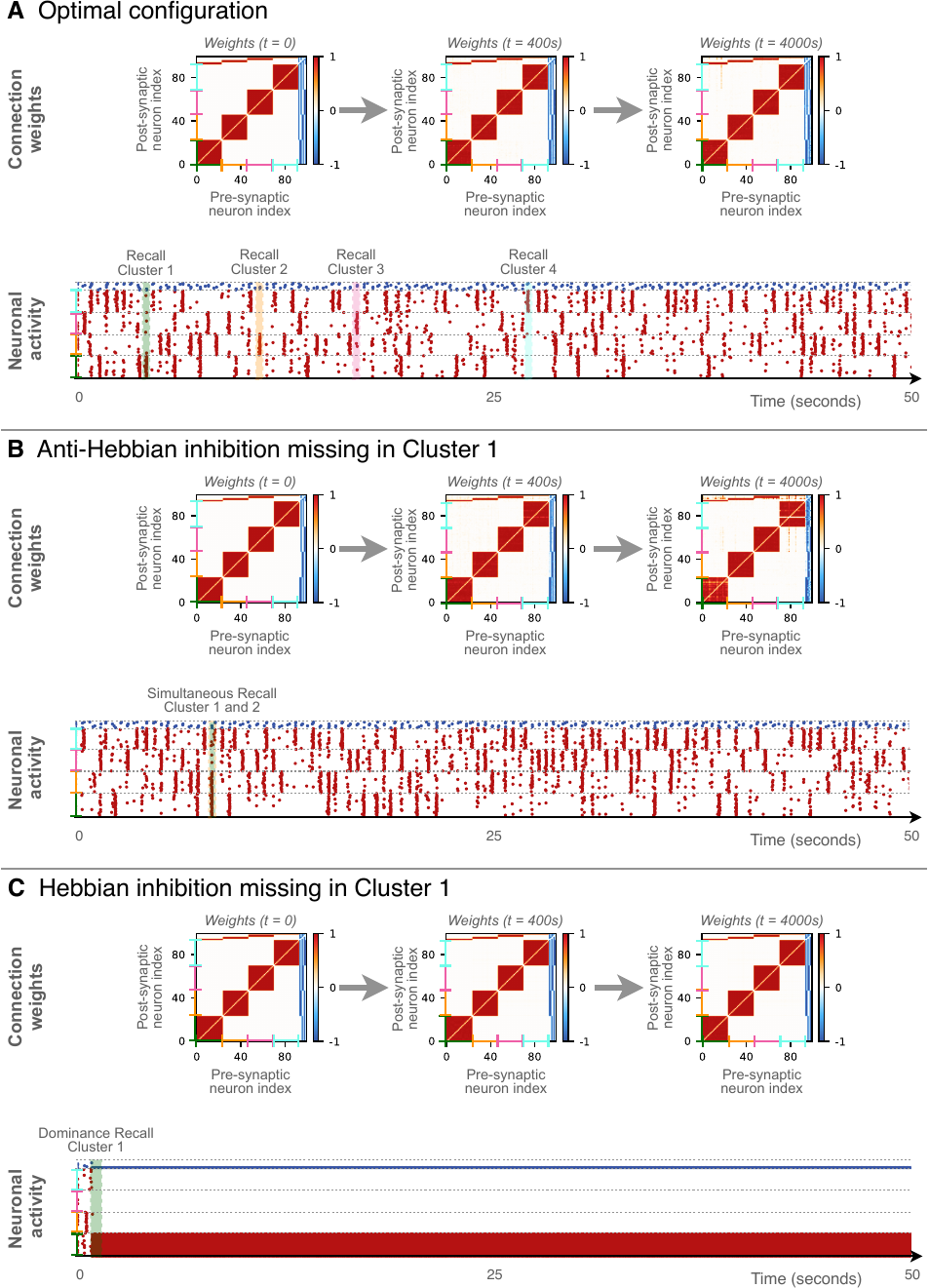}
	\caption{
			{\bf Evolution of a network initially made of four structural modules in absence of any stimulation.} 
			{\bf(A)} Network of 92 excitatory neurons and 8 inhibitory neurons (one Hebbian and one anti-Hebbian in each cluster).
			{\bf (B)} Network of 93 excitatory neurons and 7 inhibitory neurons (only an Hebbian in cluster 1).
			{\bf (C)} Network of 93 excitatory neurons and 7 inhibitory neurons (only an anti-Hebbian in cluster 1).
			In each case, we study the stability of the organization from a structural and dynamical point of view. In each panel, the connectivity matrices show the evolution of the synaptic weights and the raster plot shows the neuronal activity. The green, orange, pink and cyan brackets and shadows represent clusters 1, 2, 3 and 4. 
	}
	\label{fig:S4Appendix}
\end{figure}

The reported numerical experiments analyse the limiting cases concerning the stability of four structural modules in absence of any stimulation. In Fig.~\ref{fig:S4Appendix}A, we consider the case where each population contains only one Hebbian and one anti-Hebbian inhibitory neuron (i.e. a total of $N_I= 2 \times 4=8$ inhibitory neurons). This arrangement corresponds to the upper limit for the number of inhibitory neurons needed to maintain 4 independent memory items, represented by the red line in Fig.~\ref{fig:Figure4}B of the main text. We observe that these conditions are sufficient for each cluster to present distinct spontaneous recall in neuronal activity. 

In Fig.~\ref{fig:S4Appendix}B, we break this limit by allocating only an Hebbian inhibitory neuron to the population P$_1$. We find that even if other populations are correctly recalled, memory recall of cluster 1 can occur at similar instant to that of the other clusters. Indeed, during recall, population P$_1$ does not inhibit the activity of the other populations, letting them activate. This has a direct effect on the consolidation process, where simultaneous recall of two memories patterns tends to induce their structural merging.

In Fig.~\ref{fig:S4Appendix}C, we perform an opposite test, assigning only an anti-Hebbian inhibitory neuron to the population P$_1$. In that case, we observe a very short period of spontaneous activity with recalls from the other population. However, once population P$_1$ becomes active, it totally dominates the others, inhibiting them. These results are very similar to those observed in Fig.~\ref{fig:Figure4}B. Indeed the absence of feedback inhibition provided by Hebbian inhibitory neuron, prevents P$_1$ activity to be regulated. Ultimately, this has no direct impact on the long-term maintenance of memory items in the weight matrix since their activity is suppressed. Nevertheless, this causes issues in the processing of stored information, since the network is blocked in an abnormal state.

\vspace{1.5cm}

\clearpage
\section{Four overlapping stimuli.} 
\label{sec:supp5}

\begin{figure}[hb!]
	\centering
	\includegraphics[width=0.85\linewidth]{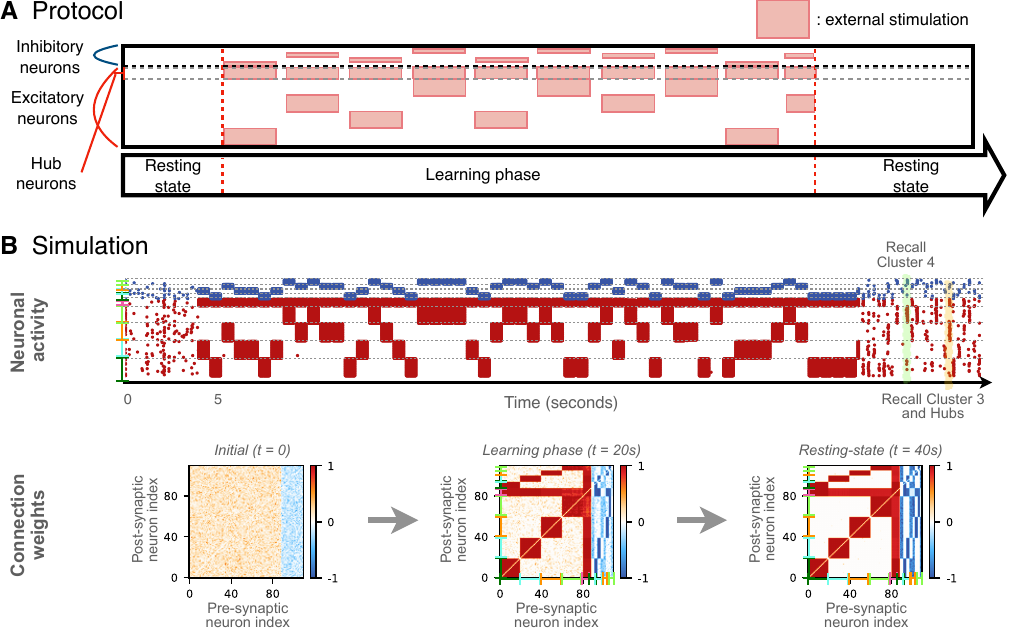}
	\caption{
		{\bf Learning of 4 overlapping stimuli.} 
		{\bf (A)} Stimulation protocol for a network of $N=110$ neurons entrained with $M=4$ stimuli that share 8 excitatory neurons. 
		{\bf (B)} Simulation and learning results. Connectivity matrices show the evolution of the synaptic weights leading to the emergence of four modules which overlap over 8 hub neurons. The raster plot shows the simulation for the three stages: initial resting phase, entrainment stage and the post-learning neuronal activity characterized by a variety of spontaneous recall events as of P$_3$ neurons with hubs (orange shadow) or P$_4$ neurons without hubs (green shadow).
	}
	\label{fig:S5Appendix}
\end{figure}

In this alternative protocol, we reproduce the experiment of Fig.~\ref{fig:Figure6} of the main text but considering $M=4$ stimuli which share 8 neurons. The results obtained are described in Fig.~\ref{fig:S5Appendix}. We obtain similar results as in main text with the formation of four modules in the connectivity matrix as expected, together with the formation of hubs which are here connected (incoming and outgoing connections) with the four clusters. Regarding the dynamics in raster plot, the resting-state activity is comparable that observed in Fig.~\ref{fig:Figure6}. We identify different types of spontaneous recalls involving one of the four clusters alone, recalls of one cluster accompanied by the hubs, or recall events of hubs alone. This experiment again highlights the richness of the different dynamics that the network can display and maintain for a certain period of time.

\vspace{1.5cm}

\clearpage
\section{Large and sparse networks.} 
\label{sec:supp6}

\begin{figure}[hb!]
	\centering
	\includegraphics[width=0.85\linewidth]{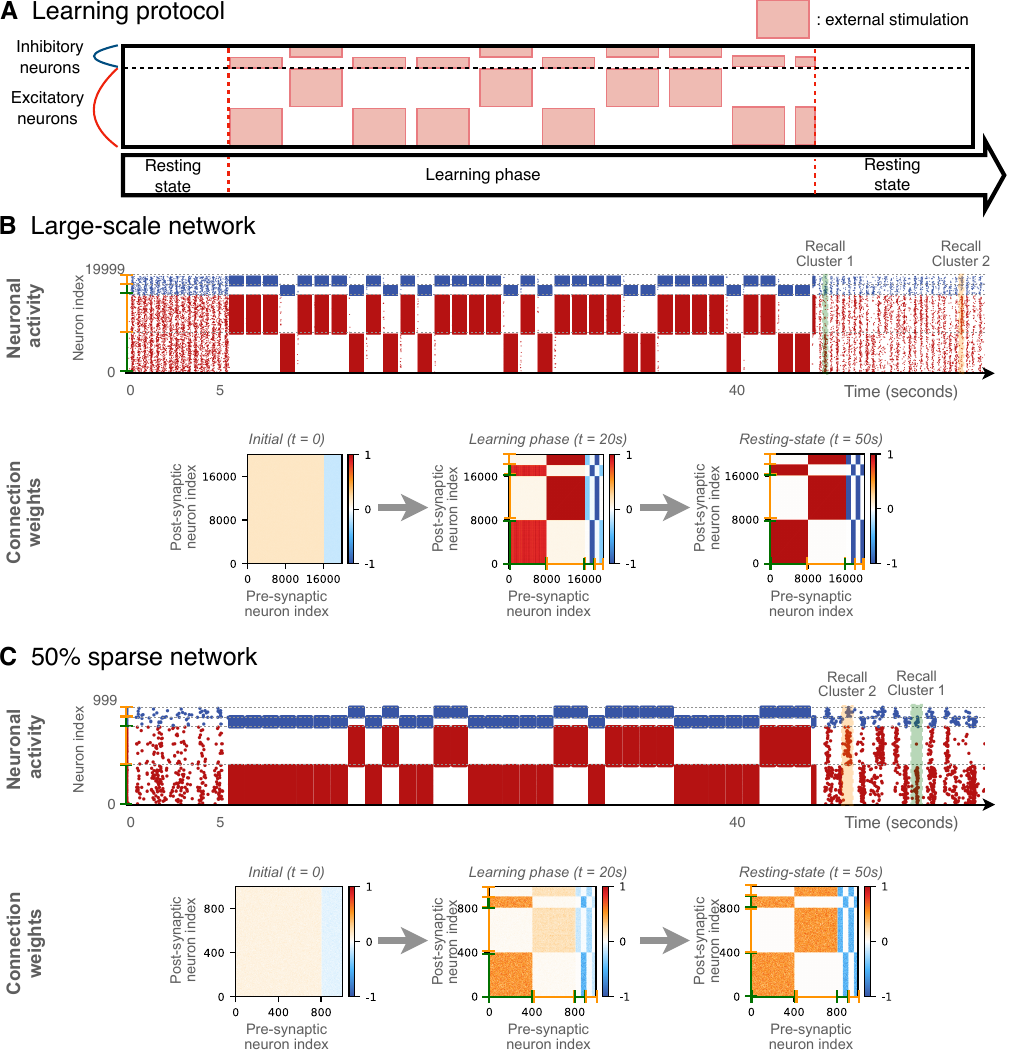}
	\caption{
		{\bf Model validation in larger and sparser networks.} 
		{\bf (A)} Stimulation protocol for a network entrained with $M=2$ stimuli. 
		{\bf (B)} Simulation and learning results for $N=20000$ neurons. Connectivity matrices show the evolution of the synaptic weights leading to the emergence of two modules. The raster plot shows the simulation for the three stages: initial resting phase, entrainment stage and the post-learning neuronal activity characterized by spontaneous recall events of P$_1$ neurons (green shadow) and P$_2$ neurons (orange shadow).
		{\bf (C)} Simulation and learning results for $N=1000$ neurons and $50\%$ sparse connection without additive noise terms. Connectivity matrices show the evolution of the synaptic weights leading to the emergence of two modules. The lighter color is due to the sparsity of the connections, but the numerical values are the same as in panel (B). The raster plot shows the simulation for the three stages: initial resting phase, entrainment stage and the post-learning neuronal activity characterized by spontaneous recall events of P$_1$ neurons (green shadow) and P$_2$ neurons (orange shadow). 
	}
	\label{fig:S6Appendix}
\end{figure}

In order to validate the stability of the model for larger network size and sparser connectivity, we apply the protocol reported in Fig.~\ref{fig:Figure1}D of the main text to networks of larger size and with randomly connected neurons, see Fig.~\ref{fig:S6Appendix}A. 
 
Firstly, we consider a network that is initially globally coupled but with random synaptic weights and additive Gaussian noise, of $N=20000$ neurons, where we still have $80\%$ excitatory and $20\%$ inhibitory neurons. The results of the experiment are reported in Fig.~\ref{fig:S6Appendix}B. 
Secondly, we consider a random Erd\"os-Reniy network composed of $N=1000$ neurons (where we still have a ratio $4:1$ for excitatory versus inhibitory neurons), with a probability of $50\%$ of possible directed connections between neurons. In this case, we omit the additive Gaussian noise to check whether the random and sparse connectivity introduces sufficient heterogeneity in the synaptic inputs of the neurons to induce an asynchronous irregular behaviour. The results of this further experiment are reported in Fig.~\ref{fig:S6Appendix}C.
 
Qualitatively, we obtain the same results as in Fig.~\ref{fig:Figure1}D, with the formation of two modular structures in the weight matrix joined to spontaneous recalls of the two different memories during the post-learning phase. The main differences are indeed observable in this regime. In Fig.~\ref{fig:S6Appendix}B, due to the size of the network it is evident that a large number of neurons fires randomly and independently, this renders more difficult to identify the spontaneous recall events. However, they still take place and the memory should be consolidated on the long term as shown in Fig.~\ref{fig:Figure5} of the main text. 

In the case of the sparse network, the activity during the resting state also presents an irregular activity of the neurons despite the dynamical behaviour is deterministic. However, also in this case, as shown in Fig.~\ref{fig:S6Appendix}C, the modular structures emerge in the weight matrix. One should notice that the orange color in the weight matrix at $t=50$ seconds is due to the fact that $50\%$ of the neurons are disconnected and do not reflect to the actual value of the weights, that have the same values as in  Fig.~\ref{fig:S6Appendix}C.

\vspace{1.5cm}

\clearpage
\section{Storing the maximum number of items.} 
\label{sec:supp7}

Here, we reproduce the experiment of Fig.~\ref{fig:Figure4}A of the main text, but considering $M=33$ stimuli. This limiting case corresponds to the maximum capacity of items that can be learned and recalled for a network of $N=100$ neurons (marked by the star symbol in Fig.~\ref{fig:Figure4}B). This implies considering a network composed of $N_{I}=66$ inhibitory neurons (with 33 anti-Hebbian and 33 Hebbian). Except these differences in the initial conditions, the network is trained as usual.

\begin{figure}[hb!]
	\centering
	\includegraphics[width=0.85\linewidth]{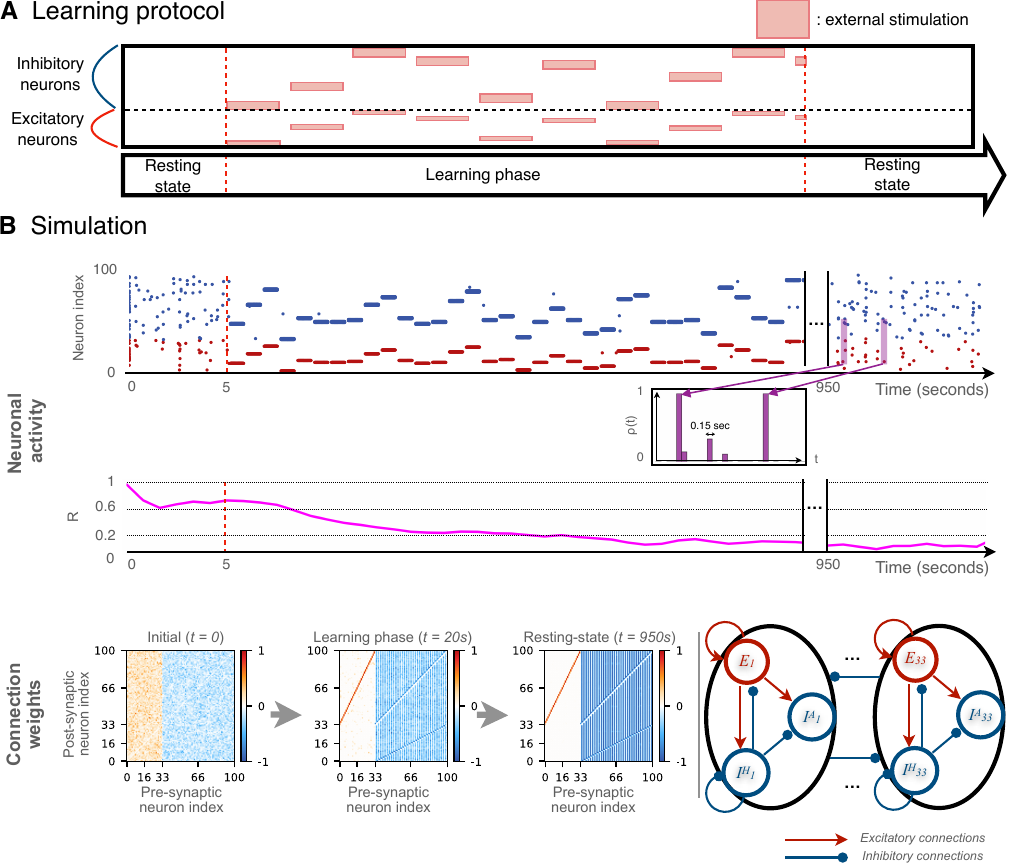}
	\caption{
		{\bf Maximum network capacity.} 
		{\bf (A)} Stimulation protocol for a network of $N=100$ neurons entrained with $M=33$ stimuli. 
		{\bf (B)} Simulation and learning results. Connectivity matrices show the evolution of the synaptic weights leading to the emergence of $33$ modules. The final configuration of the connection weights is shown schematically on the right. The raster plot shows the simulation for the three stages: initial resting phase, entrainment stage and the post-learning neuronal activity characterized by a variety of spontaneous recall events of the memories. The recalls of one memory are highlighted in purple with its normalised spike count $\rho (t)$ for bins of $0.15$ sec. The instantaneous Kuramoto order parameters $R$ of the network throughout the simulation time is displayed bellow.
	}
	\label{fig:S7Appendix}
\end{figure}

This protocol leads to the formation of $33$ memories, where each item is composed of an excitatory neuron, a Hebbian and an anti-Hebbian inhibitory neuron, as shown in the connectivity matrix and in the diagram of Fig.~\ref{fig:S7Appendix}B. Consequently, each anti-Hebbian inhibitory neuron projects to all the other clusters of neurons.
Concerning the post-learning activity, given the small size of the clusters now, it is difficult to distinguish individual memory recalls with clarity. Nevertheless, the asynchronous dynamics of the network at rest tends to confirm that all excitatory neurons have a decorrelated activity. 
To visualize this, we estimated the instantaneous Kuramoto order parameter $R$ throughout the simulation. During the post-training resting state the system is essentially desynchronized, since $R\simeq 0.1 \simeq 1/\sqrt{N}$, as expected in an asynchronous system made of $N=100$ neurons/oscillators due to the central limit theorem. On the contrary, before the training phase the system was partially synchronized, since $R>0.6$. We have also defined a normalized spike count associated to a certain memory item 
$$ \rho (t) = \frac{n^{sp}_{m}(t)}{n^{sp}(t)}$$ 
where the spike count associated to the neurons related to a given memory item $n^{sp}_{m}(t)$ is divided by the total number of spikes emitted in the network $n^{sp}(t)$. By computing $\rho(t)$ for the memory highlighted in purple in the figure, we can identify clear recalls when $\rho=1$, meaning that the spikes only come from the neurons associated to the recalled memory.

\vspace{1.5cm}

\clearpage
\section{Stability of the stored items for increasing network sizes.} 
\label{sec:supp8}

\begin{figure}[hb!]
	\centering
	\includegraphics[width=0.85\linewidth]{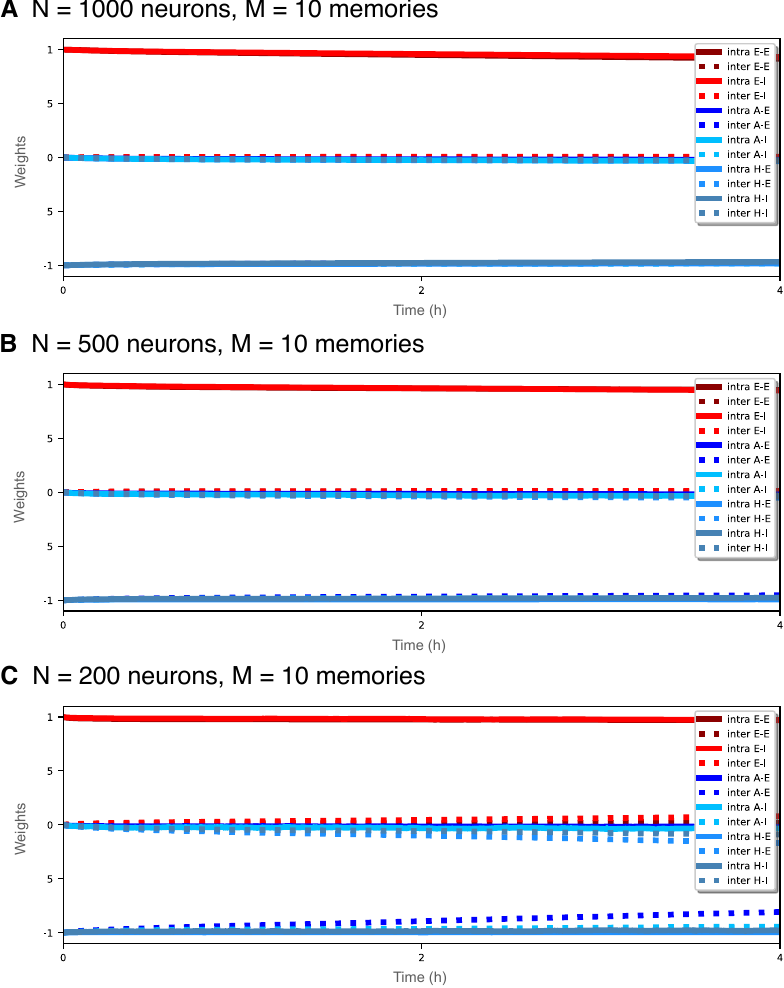}
	\caption{
		{\bf Stability of the network connections scaling up the network size.} 
		Post-learning evolution of mean intra- (solid lines) and inter- (dashed line) clusters weights for excitatory to excitatory (E-E dark red), excitatory to inhibitory (E-I red), anti-Hebbian inhibitory to excitatory (A-E dark blue), anti-Hebbian inhibitory to inhibitory (A-I cyan), Hebbian inhibitory to excitatory (H-E blue) and Hebbian inhibitory to inhibitory (H-I steel blue) connections; for {\bf (A)} $N=1000$ neurons, {\bf (B)} $N=500$ neurons, and {\bf (C)} $N=200$ neurons, all trained with $M=10$ stimuli.
	}
	\label{fig:S8Appendix}
\end{figure}

By following~\cite{yang2024}, we have analysed the stability of the intra- and inter-cluster weights for increasing system sizes and numbers of memory items (essentially restricted to $N=100$ neurons and $M=2$ memories in the main text). We have here adopted two protocols. In the first, the number of memory items is fixed to $M=10$. In the other, $M$ grows proportionally with $N$, in this way the number of neurons associated to each memory clusters remains the same. In both cases we considered three different system sizes, namely, $N=200$, $500$ and $1000$ neurons. The results obtained for $M=10$ and by varying $N$ are shown in Fig.~\ref{fig:S8Appendix}, while those corresponding to $M=N/100$ are reported in Fig.~\ref{fig:S9Appendix}.

As a general result, we observe that the intra-cluster weights essentially remain constant over time independently of $N$ for both protocols, indicating that the internal stability of the formed clusters is independent of the system size. On the other hand, for the protocol shown in Fig.~\ref{fig:S8Appendix}, we observe a tendency for the inter-clusters weights to grow/decrease over time during the post-learning phase, ultimately leading to a merging of the stored memory items. This can be explained by the fact that reducing network size, reduces the number of inhibitory neurons allocated to each memory, decreasing their stability. Indeed, even if the configurations are sufficient to guarantee coherent reactivations (see Fig.~\ref{fig:S7Appendix}), less inhibition increases the probability of incoherent spikes due to network variability. However, the rate at which the inter-cluster weights modify strongly decreases with $N$, indicating that for larger system sizes the memories remains stable for longer times. 

Finally, by considering the situation where the ratio $M/N$ is maintained constant as in~\cite{yang2024}, all intra- and inter-cluster weights remain essentially constant over 4 hours of post-learning evolution, displaying no drift in their values. By employing this second protocol we can extrapolate the memory capacity characterized in the main text for $N=100$ in Fig.~\ref{fig:Figure4}B to networks of larger size.

\vspace{1.5cm}

\clearpage
\section{Estimation of the time needed to forget a recall event. } 
\label{sec:supp9}

The rate of change of the synaptic weights depends on several factors: the current value of the weight $w_{ij}(t)$, the number of neurons spiking at a given time $t$, and the temporal precision of their spikes. Therefore, it is non-trivial to quantify the precise amplitude of the change of weights at all times. If a synapse is already at its maximum weight capacity, a potentiation (due to a recall) has almost no impact whereas the absence of recalls leads to more impactful depression (see soft bound functions in Methods, Figs.~7D and E). 

We can derive an approximation for the impact of a recall on a synaptic weight---compared to a forgetting epoch---by inspecting closer the plasticity function (see Fig.~\ref{fig:Figure7}A). For uncorrelated spikes during an asynchronous irregular firing epoch (i.e. for $|\Delta t| > 0.5$), the weights depress proportionally to $\Lambda(\Delta t) \approx -0.1$ due to the forgetting term. On the other hand, if the spikes are fully synchronized (i.e. $\Delta t=0$), their increase is proportional to $\Lambda(\Delta t) = A_+-A_- -f = 2.347 - f$. Thus, by considering an average weight of $w_{ij} = 0.5$ (where  $\tanh(\lambda(1 - w_{ij}))$ = $\tanh(\lambda w_{ij})$  $\approx$ $1$) Eq.~10 becomes:
\begin{equation}
	\label{eq:3_20}
	\frac{[w_{ij}(t^+) - w_{ij}(t^-)]}{\gamma_l} \simeq
	\begin{cases}
		{2.347 -f,}&{\text{for synchronized spikes during a recall}} \enskip , \\
		{-f ,}&{\text{for uncorrelated spikes during AI state} \enskip .}
	\end{cases}
\end{equation}

The ratio of these two (absolute) values for potentiation and depression for $M$ stimuli and $f = f_0/M$ is: $\frac{2.347}{f_0} M - 1 = 11.735 M - 1$, where $f_0=0.2$, and it gives an estimate of the number of uncorrelated spikes that would be required to lose the acquired potentiation. The number of these spikes grows linearly with $M$. By considering an average firing rate of $2$ Hz for the uncorrelated firings, this implies that a period of $\simeq 11$ seconds would be required to forget the contribution of a single recall event for $M=2$, that will grow to $\simeq 58$ seconds for $M=10$. The prolongation of this time window proportionally to $M$ will allow to all the stored memory items to be randomly recalled before they are forgotten.

\end{document}